\documentclass[preprint,12pt]{elsarticle}

\usepackage{graphicx}
\usepackage{epsfig}

\usepackage{amssymb}

\journal{Physica D}

\begin{document}

\begin{frontmatter}



\title{Stochastic (in)stability of synchronisation of oscillators on networks}


\author[label1]{Mathew L. Zuparic}
\ead{mathew.zuparic@dsto.defence.gov.au}
\author[label1]{Alexander C. Kalloniatis}
\ead{alexander.kalloniatis@dsto.defence.gov.au}
\address[label1]{Defence Science and Technology Organisation, Canberra,
ACT 2600, Australia}

\begin{abstract}
We consider the influence of correlated noise on the stability of synchronisation of 
oscillators on a general network using the Kuramoto model for coupled
phases $\theta_i$. Near the fixed point $\theta_i \approx \theta_j \
\forall i,j$ the impact of the noise is 
analysed through the Fokker-Planck equation. 
We deem the stochastic system to be `weakly unstable' if the 
Mean First Passage Time for the system to drift
outside the fixed point basin of attraction is less than 
the time for which the noise is sustained. We argue that a Mean First Passage Time, computed near the 
phase synchronised fixed point, gives a useful lower bound on the tolerance of the system to noise.
Applying the saddle point approximation, we
analytically derive general thresholds for the noise parameters for weak stochastic stability.
We illustrate this by numerically solving the full Kuramoto model in
the presence of noise for an example complex network.
\end{abstract}


\end{frontmatter}


\section{Introduction}
\label{intro}
The spontaneous generation of ordered or patterned behaviour is a significant property of 
dynamical systems of many nonlinearly coupled heterogeneous entities. This behaviour is 
relevant to a vast array of real world systems: physical, chemical, biological, ecological
and social. Where the dynamics of the component level entities have cyclic characteristics,
the model of coupled phase oscillators on a network is a compact 
mathematical representation offering much insight. 
The Kuramoto model \cite{Kura} is a simple example of this,
exhibiting transition from incoherence to phase synchrony as a single coupling constant
is varied from small to large values. In this paper we explore how thresholds for
the strength of noise applied
for a finite period of time to the system close to the synchronised state can 
be determined such that stability is maintained.

The Kuramoto model \cite{Kura,Strog00} involves one-dimensional phase
oscillators described by angles $\theta_i$ coupled on a network described by an 
undirected graph $G$ of size $N$. The model is given by the first order 
differential equation,
\begin{eqnarray}
\dot{\theta}_i = \omega_i - 
\frac{K}{N} \sum^N_{j=1} A_{ij} \sin(\theta_i-\theta_j) ,& 1 \le i \le N.
\label{KurMod}
\end{eqnarray}
Here $A_{ij}$ is the adjacency matrix encoding the graph structure, $\omega_i$ 
are intrinsic frequencies
of the oscillators, usually drawn from some statistical distribution, and $K$ 
is a global coupling controlling the strength of mutual adjustment between
adjacent oscillators.  At zero coupling the system behaves
incoherently. Increasing the coupling drives the bulk of
oscillators to lock into a single core that undergoes
harmonic motion according to the average of the frequencies, 
while others undergo random motion about the unit circle. 
At lower coupling values phase {\it velocities} can synchronise, 
$\dot{\theta_i}\approx\dot{\theta_j}$, and at stronger coupling 
the {\it phases} themselves synchronise, $\theta_i\approx\theta_j$.
We are interested in the regime where the latter applies to all nodes,
and will refer to this as `phase synchronisation'.
Kuramoto \cite{Kura} considered the complete ($A_{ij}=1$) 
infinite ($N\rightarrow \infty$)
case, which can be solved by fluid dynamics techniques.
Stability and noise have been separately studied with such methods in
\cite{Sak88,MirStrog91}. Unfortunately, these cannot be
generalised to non-complete finite graphs, in
which we are interested. Similarly, it is difficult to draw general conclusions 
from numerical studies of specific graphs such as \cite{GomGard07}.

Noise is invoked either because the fine-structure of component entities cannot
be fully encoded in deterministic equations such as Eq.(\ref{KurMod}), or 
to undermine/encourage known deterministic behaviours. One is interested then in the threshold for 
tolerance to noise either to understand the limitations of a deterministic model of a system or the capacity of the 
known deterministic system to cope with imposed noise.
In \cite{Sakaguchi}, Sakaguchi applied stochastic flucuations to the intrinsic frequencies in Eq.(\ref{KurMod}) and 
obtained an equivalent result to Kuramoto in \cite{Kura} in the complete infinite case. Park, Kim and Ryu \cite{ParkKim96,Kim97} have numerically examined noise in the 
equal frequencies and complete network Kuramoto model
(using an extra `pinning term') but, again, their insights cannot be generalised.
Analytic approaches to deterministic (Lyapunov) stability of synchronisation, such as 
that of Pecora and Carroll \cite{PecCar98}, are welcoming 
because they open the door to analysis of general graphs and frequencies.
The non-polynomial interactions in Eq.(\ref{KurMod})
mean that the method of \cite{PecCar98} only works close to phase synchrony
\cite{Kall10}. Noise can nevertheless be examined in the neighborhood of the fixed point basin of attraction.
It is known that,  except for special cases of stochastic
systems that are `classically stable' \cite{Schuss10},
noise leads to drift outside some bounded region if an arbitrary amount of
time is permitted. However, for many biological and social systems, 
the finite life of entities mean that noise need/may not be sustained indefinitely.
There is value then in a criterion for `weak stochastic stability'
for the Kuramoto model, so that stability of phase synchronisation can be
guaranteed within a {\it finite period of application of noise}.

We emphasise that Eq.(\ref{KurMod}) \textit{combined with noise} is 
a stochastic generalisation of the Kuramoto model. 
Noise can be injected through a combination of fluctuations on the coupling, 
the frequencies, and the network topology.   
It is known that near phase synchrony a certain decoupling of the dynamics into collective modes occurs
\cite{Kall10,McGMenz08}. We therefore adopt the strategy of applying noise to these collective modes.
We solve for the stationary probability distributions for these modes from the Fokker-Planck equation. 
By then computing
the Mean First Passage Time (MFPT) for the modes to cross an effective
boundary of the basin of attraction for the phase synchronised fixed point
we may analytically derive thresholds for various parameters, thereby
guaranteeing weak stochastic stability. 

By basing ourselves in an approximation near the phase synchronised fixed point, our results do not include effects of recoupling by 
the modes as noise pushes them outside their basin of attraction, which in turn can lead to quite exotic {\it stabilising} effects in the presence of noise.
This makes our analytical determination of a Mean First Passage Time a strong under-estimate of the full non-linear result. However, this provides a useful lower bound on
tolerances of the system to noise-driven instability. This result enables 
thresholds to noise to be computed for general graphs and frequencies. 
The focus of our paper is what can be computed analytically, however we illustrate our results - including the nonlinear effects our analytic approach omits - by numerically solving
the full Kuramoto system for a specific non-trivial graph.

We formulate the model in the next section, and then
discuss our solution method for the Fokker-Planck system; we show here an example complete 
numerical solution exhibiting noise generated instability. 
We next state the results for the stationary probability distribution functions 
in preparation for characterising the stochastic stability in terms of the MFPT and show how
general stability thresholds in the noise parameters can be derived.
Subsequently, with a numerical
solution of the full model we illustrate how, using the MFPT computed near phase synchrony, stochastic instability
can be avoided. We discuss implications of our results for
general graphs and look ahead to future work. Details of some derivations are relegated to three appendices.

\section{Stochastic equations close to synchrony}
\label{formulation}
\subsection{Laplacian decomposition of deterministic case}
As mentioned in the introduction, close to phase synchrony the dynamics decouples into independent 
behaviours of certain collective modes \cite{Kall10,McGMenz08}. This decoupling is based on a decomposition of
the network into eigenvectors of the {\it graph Laplacian}.
For a connected unweighted undirected graph $G$ the Laplacian is defined as
\cite{Boll98}
\begin{equation}
L_{ij} = d_i \delta_{ij} - A_{ij},
\label{LaplDef}
\end{equation}
with $d_i$ the degree of node $i$. 
The spectrum of Laplacian eigenvalues $\lambda_i$ is positive semi-definite,
as explained by Bollob{\'a}s in \cite{Boll98}. The corresponding system of orthonormal eigenvectors ${\vec \nu}^{(r)}$ satisfy 
\begin{eqnarray}
\sum^{N}_{j=1} L_{ij} \nu^{(r)}_j = \lambda_r \nu^{(r)}_i, && \sum^N_{i=1}\nu^{(r)}_i  \nu^{(s)}_i = \delta_{rs} . 
\nonumber 
\end{eqnarray}
Most immediately useful here is the property that the lowest eigenvalue vanishes, $\lambda_0=0$. 
For a connected graph the first non-zero eigenvalue $\lambda_1$ is known as the 
algebraic connectivity \cite{Fied73}.
Graphs with $\lambda_1<1$ are easily disconnected by removal of a small number of links.
\begin{figure}[htb]
\centering
\includegraphics[width=120mm]{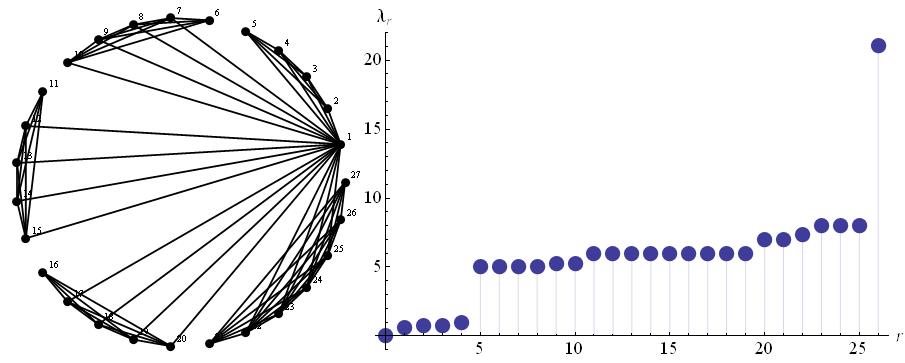}
\caption{An `RB'-type network represented in a circular embedding with nodes numbered anti-clockwise
shown in the upper panel (left) and its associated Laplacian spectrum (right).} 
\label{RB-Graph-LaplSpectrum}
\label{firstfig}
\end{figure}

An example of a connected graph of $N=27$ nodes, 
a variation of a graph due to Ravasz and Barab{\'a}si (RB) \cite{RavBar03}, and 
its related Laplacian spectrum are shown in Fig.\ref{RB-Graph-LaplSpectrum}; 
we shall later use this `RB' graph to illustrate the properties we derive in this paper. In
particular, we observe in the spectrum of Fig.\ref{RB-Graph-LaplSpectrum} the single zero 
eigenvalue, a sequence of four eigenvalues less than one, a broad set between values five and ten, and
then the jump at $N=26$ with $\lambda_{26}= 21.08$. 
That there are four eigenvalues less than one, close to $\lambda_1$, is indicative
of the hierarchy and symmetry of the graph; the circular embedding
in Fig.\ref{RB-Graph-LaplSpectrum} reveals a number of ways of disconnecting
the graph with four link removals. The proportions of
eigenvalues less than one and, respectively, greater than one will vary for
graphs of different classes. 

\begin{figure}[htb]
\centering
\includegraphics[width=130mm,angle=0]{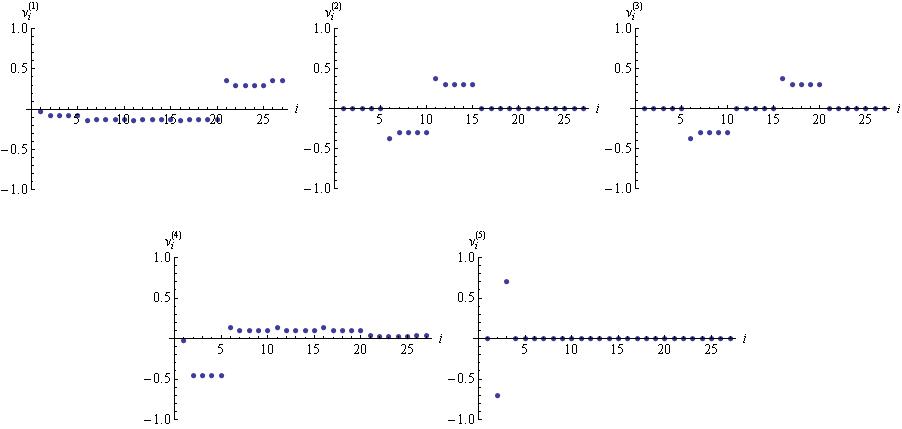}
\caption{Plots of the values of the components of the $r=1,\dots,5$ normalised Laplacian eigenvectors for
the RB graph.} 
\label{RBevector}
\end{figure}

In considering how noise can be applied to a network in this context, it is important to explain the 
network interpretation of the
Laplacian eigenvectors $\nu^{(r)}_i$. They represent `weighted' {\it sub-graphs} of $G$: 
the component $\nu^{(r)}_i$ is the (not necessarily positive) numerical weight with which a node $i$ is represented
in the {\it r-th} sub-graph. 
For the RB graph in Fig.\ref{RB-Graph-LaplSpectrum}, for example, the 
eigenvector corresponding to the third eigenvalue $\lambda_3=0.764$ is
shown, alongside the eigenvectors for $r=1,\dots, 5$, in Fig.\ref{RBevector}.
Thus $\nu^{(3)}$ describes a subgraph consisting of the nodes 
$i=6,\dots,10,16\dots,20$ 
with both positive and negative
weights applied. We may further say that within the sub-graph, the nodes $i=6,\dots,10$ 
and $i=16,\dots,20$ form finer sub-structures that are internally
correlated but anti-correlated with respect to each other.
Such sub-graphs are 
the network analogue of the combinations of atomic bonds that give rise to 
phonons as collective excitations in crystal lattices.
In Fig.\ref{RBevector} we observe quite different characteristics in the structures
of the lowest lying eigenvectors
in terms of the proportion of positive, negative and zero values. 

The Laplacian is, in fact, a Laplace-Beltrami operator on the manifold described by $G$.
To this extent it reflects properties well-known 
in condensed matter systems: the eigenvalues $\lambda_r$ are a `momentum-squared' and 
their corresponding eigenvectors act like `probes' on the sub-graphs. The eigenvectors
corresponding to low-lying eigenvalues $r\sim 1$ involve larger scale sub-graphs
of $G$ while those for higher values $r\sim N$ reflect finer structures in the graph.
This `probing' of sub-structures will vary according to the precision with which 
one may quantify or manipulate properties of 
nodes of the network. If the network represents a set of physical or chemical entities one 
may have a considerably precise means of measuring or altering properties of the nodes. For example, 
in the case of condensed matter systems one is typically able to excite individual frequency 
modes of phonons.
Contrastingly, if the network is a biological or social system the means of 
measuring or manipulating the properties may be more limited. Either way, it is important that 
the probing of sets of nodes may be distinguished in relative strength and in sign, that 
it may be applied positively or negatively at different parts of the network according to 
the structure of an eigenvector.
We discuss this further below in the context of applying noise.

We now expand the phase angles of the Kuramoto model in the Laplacian
basis of eigenmodes,
\begin{eqnarray}
\theta_i(t) = \sum^{N-1}_{r=0} x_r(t) \nu^{(r)}_i, && x_r(t) =   \sum^{N}_{i=1} \theta_i(t) \nu^{(r)}_i.
\label{mode-exp}
\end{eqnarray}
We refer to the $x_r$ as the Laplacian `modes' of the Kuramoto system.
Treating $\omega_i$ as the components to an 
$N$-dimensional vector, $\vec \omega$, with $\bar\omega$ the average over the 
frequency distribution, 
we can form the scalar product of the frequency vector and the Laplacian eigenvectors
\begin{eqnarray}
\omega^{(r)}=\vec{\omega }\cdot \vec{\nu}^{(r)}.
\nonumber 
\end{eqnarray}
We now consider the system close to a point of phase
synchronisation ($\theta_i\approx \theta_j \ \forall i,j$),
\begin{equation}
\dot{\theta_i}  \approx  \omega_i - \sigma \sum^N_{j=1} L_{ij} \theta_j,
\label{motion2}
\end{equation}
where from this point we summarise the coupling as $\sigma=K/N$. 
Inserting the decomposition Eq.(\ref{mode-exp}) into the truncated equation of motion 
Eq.(\ref{motion2}), 
we can extract equations for the eigenmodes $r\neq 0$,
\begin{equation}
\dot{x}_r(t) = \omega^{(r)} - \sigma  \lambda_r x_r(t)  
\label{logistic}
\end{equation}
while the zero mode of Eq.(\ref{motion2}) gives $x_0(t)=\sqrt{N}(t-t')\bar{\omega}$
at time $t$ after initial time $t'$. We see that, indeed, 
to this order, the modes $r$ are decoupled.
The exact solution to Eq.(\ref{logistic}) is:
\begin{equation}
x_r(t) = x_r ' e^{- \sigma \lambda_r (t-t')} + \frac{\omega^{(r)}}{\sigma \lambda_r}
\left( 1 - e^{- \sigma \lambda_r (t-t')}  \right)
\label{detersol}
\end{equation}
where $x_r'=x_r(t')$ represents the initial configuration.
Taking the large $t$ limit of this shows that there is a fixed point 
\begin{equation}
x_r^{*}=\frac{\omega^{(r)}}{\sigma\lambda_r}
\label{fixedpointdef}
\end{equation}
whose stability is determined by
the Lyapunov exponents $-\sigma\lambda_r$ in Eq.(\ref{detersol}). They do not change sign
due to the semi-definiteness of the Laplacian spectrum \cite{Boll98}. 
Thus phase synchrony is stable in every direction of trajectory space \cite{Kall10}
(whose axes can be labelled by the modes $r$): 
fluctuations on $\theta_i=\theta_j \ \forall \ i,j$ are small,
\begin{equation}
|x_r(t)|\ll 1 \ \forall t
\label{stabcrit}
\end{equation}
as long as the condition is true at the initial time $t=t'$, and 
\begin{equation}
|x_r^{*}| \equiv \left|\frac{\omega^{(r)}}{\sigma\lambda_r}\right| \ll 1
\label{fixedpoint}
\end{equation}
so that the approximations leading to
Eq.(\ref{motion2}) are valid over all times. 

\subsection{Bounding the basin of attraction}
The $x_r^*$ define the origin of 
a basin of attraction in trajectory space whose curvature in 
every direction of the origin is concave since the $\sigma \lambda_r$ are all positive.
The boundary of the basin
is determined by the range in which $\sin( \Delta\theta) - \Delta\theta \approx 0$, 
with $\Delta\theta$ the difference between any two phases. This leads to
$\Delta\theta\approx 0.5\equiv \Delta\theta_c$. The quantity $\Delta\theta_c$ is the threshold difference 
in phases for which the linear approximation to sine is
deemed valid. To convert this into a general boundary on $x_r$
is difficult. 

Some authors have, effectively, derived bounds on norms over the modes \cite{Jad04,DorfBull11} in order to 
give the critical coupling in terms of some aggregate
property of the frequency distribution - in the spirit of Kuramoto's analytical result for the 
critical coupling of the $N\rightarrow\infty$ complete graph \cite{Kura}. 

Our experience with numerical solutions of systems with 20-30 nodes \cite{Kall10} is that $x_r=\pm 1$
is a sufficient indicator of the boundary. We shall see this also in the specific example
of the RB graph we use in this work. 
In Appendix A we show, based on the structure of underlying Laplacian eigenvectors and the scaling of their components, 
how this heuristic choice of the boundary scales with the 
size of the graph $G$. In the interests of brevity we simply state the basic result here.

For graphs where the lowest non-zero eigenvalue is of the order $N$, $\lambda_1\sim N$, the graph is highly connected, close to a complete graph. The corresponding 
low-lying eigenvectors
are then highly `localised', namely have non-zero components focused in two nodes:
$\nu^{(r)}_i =\pm 1/\sqrt{2}$ for two values of $i$. In that case, the boundary of the basin in $x_r$ can be taken as
\begin{eqnarray}
x_r =\pm \sqrt{2}.
\end{eqnarray}

At another extreme, a large graph which is easily disconnected into two nearly complete graph components of size $N/2$ by a few link removals, will have an
eigenvector $\nu^{(1)}_i$ which will look like a step function, going from values $-1/\sqrt{N}$ to $+1/\sqrt{N}$ at the nodes that form the bridge between the two parts. In that case the boundary of the basin in $x_r$ scales with $\sqrt{N}$:
\begin{eqnarray}
x_r =\pm \sqrt{N}.
\end{eqnarray}

In between these two extremes are graphs whose sub-structures do not scale uniformly with $N$: for example, reproducing the structure of the RB graph (which is designed to exhibit hierarchy and thus be scale-free) for larger $N$. For these the simple scaling arguments
we give in Appendix A do not work. Given that at the extremes the size of the basin in $x_r$ increases, it follows that 
maintaining a heuristic boundary at $x_r=\pm 1$ is an under-estimate. Since we are looking for thresholds for stability under noise,
it is reasonable we remain with this choice in the following.

The independence of Laplacian modes, which survives considerably far from the regime 
where the `smallness' criterion of 
Eqs.(\ref{stabcrit}) and (\ref{fixedpoint}) applies \cite{McGMenz08}, 
leads to the following simple intuitive picture. 
The Laplacian eigenmodes with largest values of
$x_r^{*}$ indicate whether the system
can phase synchronise. This quantity brings together the three key elements of the
deterministic system: the frequencies, the network topology (through the 
Laplacian eigenvalues and eigenvectors) and the
coupling constant. There are a number of ways the elements may combine to satisfy
Eq.(\ref{fixedpoint}): sufficiently strong coupling, sufficiently narrow frequency range,
and sufficiently large $\lambda_r$, which amounts to sufficiently strong
connectivity.
Thus the mode most susceptible to disruption from synchronicity
corresponds to that $r$ for which $x_r^{*}$ is closest to one,
which is not necessarily that corresponding to $r=1$. 

We obtain from Eq.(\ref{fixedpoint}) a {\it weak} type of criterion for synchronisation
\cite{Jad04},
$\sigma > |\omega^{(r)}|/\lambda_r$ $\forall r.$  
But because we
cannot be more precise about the basin boundary we cannot, at this order, 
extract a sharp threshold of
coupling which distinguishes synchrony from incoherence as distinct phases.
Second order fluctuations offer some scope for overcoming this
\cite{Kall10}. However, noise induced instabilities are our 
concern in this paper. Specifically, we shall pose the question: {\it if the 
deterministic system is stable, according to Eq.(\ref{fixedpoint}), what parameter
choices in the noise will render the stochastic system unstable?}

\subsection{Applying noise: Stochastic system in Laplacian decomposition}
\label{Applying noise}
The well-known logistic equation $\dot{x}=x(1-x)$ for population dynamics can be brought
to the form Eq.(\ref{logistic}) after a change of variables.
This is our clue
for examining the system in the presence of stochastic fluctuations: we may exploit
a significant body of work on noisy population dynamics. 
Specifically, in \cite{2} and references therein, Go\'{r}a considered a 
generalisation of the logistic equation 
by perturbation with two correlated Gaussian White Noise (GWN) terms, 
$\Gamma^{(a)}$ and $\Gamma^{(m)}$; the superscripts $a$ and $m$ stand for additive and 
multiplicative respectively. 

We perform a similar analysis by letting the deterministic system 
reach a state close to synchrony (Eq.(\ref{motion2})). At time $t'$ we apply a stochastic 
disturbance to the system for a finite period of time $\kappa_T$. 
We wish to apply noise both additively, on top of the static 
but otherwise random frequencies at each node, and multiplicatively - 
namely to the coupling strength of the links between each node.
Hence for each graph node $i$ we apply the following \textit{additive} 
weighted sum $\Lambda_i(t)$ of random time dependent noises, 
\begin{equation}
\Lambda_i (t)= \sum^{N-1}_{r=0} \gamma^{(r)}_2  \nu^{(r)}_i\Gamma^{(a)}_r(t)S(t),
\label{noisefluct1}
\end{equation}
where the Heaviside step functions, $S(t)\equiv \Theta(t-t') \Theta(\kappa_T-t)$, 
act to switch the noise on and off.
Similarly, for each graph link $(i,j)$ we apply the following \textit{multiplicative} 
weighted sum, $\Lambda_{ij}(t)$, of random time dependent noises,
\begin{equation}
\Lambda_{ij}(t) = -\sum^{N-1}_{r=0} \gamma^{(r)}_1 \nu^{(r)}_i \nu^{(r)}_j\Gamma^{(m)}_r(t) 
S(t),
\label{noisefluct2}
\end{equation}
where $\{\gamma^{(r)}_1, \gamma^{(r)}_2 \} \in \mathbb{R}$. 

For the full Kuramoto system, applying the noise Eqs.(\ref{noisefluct1},\ref{noisefluct2}) 
leads to the following Langevin-like equations \cite{Risken}:
\begin{eqnarray}
&&\dot{\theta_i} = \omega_i + \Lambda_i (t) -  \frac{K}{N} \sum^N_{j=1} A_{ij} \sin(\theta_i-\theta_j)
- \sum^N_{j=1} \Lambda_{ij}(t)\theta_j(t). 
\label{fullKura-noise}
\end{eqnarray}

$\Lambda_i (t)$ in Eq.(\ref{fullKura-noise}) represents introducing noise on 
the natural frequencies, $\omega_i$, of each graph node,
while $\Lambda_{ij} (t)$  represents
noise on the interaction between nodes $i$ and $j$. This form means that nodes $i,j$ for which
$A_{ij}=0$ may nonetheless have noise applied across them. For a non-complete graph there is no natural structure for applying noise multiplicatively other than the graph itself,
via $\sigma\rightarrow \sigma+\Gamma$ for all $i$ (as in \cite{ParkKim96,Kim97}). Our prescription is, therefore, more general than applying noise 
uniformly to the existing connections in the static graph given by $A_{ij}$. However, it is by no means the most general type of noise that could be applied
since there are only $N$ degrees of freedom (the values of $\gamma_1^{(r)}$) for $N^2$ possible entries in a general
matrix $\Lambda_{ij}$. Depending on the instantiation of the network, specific non-factoriseable forms for $\Lambda_{ij}$ may arise organically.
The advantage of our form is that the Fokker-Planck equation is factoriseable, and therefore analytically tractable, permitting us to
make some definite statements of the impact of noise on stability of the fixed point. 

Note also that the multiplicative noise in Eq.(\ref{fullKura-noise}) is linear in the phase $\theta_i$, implying that for 
large phases the multiplicative noise will dominate
over the bounded sine interaction. This is certainly quite strong, but ultimately
we are only interested in the 
behaviour of the system in the vicinity of phase synchrony so that the dynamics beyond the 
loss of synchrony will not be studied in any detail.
Near phase synchrony, therefore, the system can be expressed either in terms of the
phase angles or eigenmodes:
\begin{eqnarray}
&&\dot{\theta_i} = \omega_i + \Lambda_i (t)-  \sum^N_{j=1}\left\{ \sigma L_{ij} 
+\Lambda_{ij}(t)\right\}\theta_j(t), \label{gora?}\\
&&\dot{x}_r(t) = \omega^{(r)} + \gamma^{(r)}_2 \Gamma^{(a)}_r (t) S(t) -  
\left\{\sigma  \lambda_r + \gamma^{(r)}_1 \Gamma^{(m)}_r (t) S(t)\right\} x_r(t) .
\label{gora}
\end{eqnarray}

The additive and multiplicative GWN terms are themselves given in terms of
uncorrelated GWN functions $\Gamma^{(1)}_r$ and $\Gamma^{(2)}_r$ through
\begin{eqnarray}
\Gamma^{(a)}_r(t) = c_r \Gamma^{(1)}_r(t) +  \sqrt{1-c_r^2} \Gamma^{(2)}_r(t), 
& \Gamma^{(m)}_r(t) = \Gamma^{(1)}_r (t),
\label{GWN}
\end{eqnarray}
where the following expectation values encode the absence of correlation between
$\Gamma^{(1)}_r$ and $\Gamma^{(2)}_r$:
\begin{eqnarray}
\langle \Gamma^{(1)}_r (t)  \rangle = \langle  \Gamma^{(2)}_r(t) \rangle = 
\langle \Gamma^{(1)}_r (t) \Gamma^{(2)}_r(t') \rangle = 0 ,
\nonumber \\
\langle \Gamma^{(1)}_{r_1}(t) \Gamma^{(1)}_{r_2} (t') \rangle =  \langle  \Gamma^{(2)}_{r_1}(t) 
\Gamma^{(2)}_{r_2} (t') \rangle = \Omega \delta_{r_1 r_2} \delta(t-t').
\nonumber 
\end{eqnarray}

The quantity $\Omega \in \mathbb{R}_+$ is the diffusion constant. 
(As $\Omega$ approaches zero we obtain the usual deterministic results.) 
Hence $\Gamma^{(a)}_r$ and $\Gamma^{(m)}_r$ are 
correlated in the following way,
\begin{eqnarray}
\langle \Gamma^{(a)}_{r_1} (t) \Gamma^{(m)}_{r_2} (t') \rangle = c_{r_1} 
\Omega \delta_{r_1 r_2}\delta (t-t').
\label{corr}
\end{eqnarray}
We refer to the constant $-1\le c_r \le 1$ as the \textit{correlation parameter} 
as it provides an indication of the mutual correlation between the 
noises applied to the natural frequencies and link interactions. 

\subsection{Solving the Langevin equations}
Since Eq.(\ref{gora}) is linear in $x_r$ we can apply standard techniques to obtain the solution 
\begin{eqnarray}
x_r(t) = \int^{t_1=t}_{t_1=t'}dt_1 Y(t_1) e^{\int^{\tau=t_1}_{\tau=t'}d \tau Z(\tau)}+ x^{'}_r  e^{-\int^{\tau=t}_{\tau=t'}d \tau Z(\tau)},
\label{sollang}\end{eqnarray}
where
\begin{eqnarray*}
Y(t) =\omega^{(r)}+\gamma^{(r)}_2 \Gamma^{(a)}_r (t) S(t) , && Z(t) = 
\sigma \lambda_r + \gamma^{(r)}_1 \Gamma^{(m)}_r (t) S(t). 
\end{eqnarray*}
For short time solutions we can obtain the following approximation to Eq.(\ref{sollang}),
\begin{eqnarray}
x_r(t=t'+\epsilon) \approx x^{'}_r 
e^{-\left\{ \sigma \lambda_r+ \gamma^{(r)}_1 \mathbb{E}\left[ \left. \Gamma^{(1)}_r(t) \right| t \in (t',t'+\epsilon)\right] \right\} \epsilon},
\label{shortlang}
\end{eqnarray}
where $\epsilon \ll 1$. Here $\mathbb{E}(A|B)$ means the expectation value of $A$ given condition $B$. We see that the deterministic Lypunov behaviour,
given through $-\sigma\lambda_r$, picks up a stochastic contribution; see also \cite{KloPlat99}.

\subsection{Classical stochastic stability}
\label{classstoch}
In \cite{Schuss10}, stochastic stability is defined as 
the property that a stochastic system will remain in a neighborhood of a fixed point for all time. 
We can identify such a case of the near-synchronised
Kuramoto model with the following parameter assignments: 
$c_r=1$, $\gamma^{(r)}_1=\mp \gamma^{(r)}$, 
$\gamma^{(r)}_2=\mp \frac{\omega^{(r)}}{\sigma \lambda_r} \gamma^{(r)}$ 
and $y_r = \omega^{(r)}-\sigma \lambda_r x_r$. This gives the Langevin system
\begin{equation}
\dot{y}_r = - \sigma \lambda_r y_r \pm \gamma^{(r)} y_r \Gamma^{(1)}_r.
\label{class-stab}
\end{equation}
Following \cite{Schuss10} for Eq.(\ref{class-stab})
(for more details refer to \ref{stabby}) we find that with 
\begin{equation}
\left( \gamma^{(r)} \right)^2 < 2 \sigma \lambda_r
\label{stabreq}
\end{equation}
the system indeed remains close to the fixed point for all time. Note that later, when we numerically solve
Eq.(\ref{fullKura-noise}), we shall see solutions that stay close to the fixed point for all time. These are due to interactions inaccessible to the 
approximation to first order in fluctuations. Thus Eq.(\ref{stabreq}) leads to stochastically stability within the bounds of first order fluctuations.

We refer to the requirement that the system stay in the vicinity of
the fixed point for $\kappa_T\rightarrow\infty$ as {\it classical stochastic stability},
an unnecessarily restrictive condition
for many real world systems with living entities. 
We instead adopt a more flexible definition of 
stochastic stability based on the \textit{Mean First Passage Time} and 
the finite time $\kappa_T$ for which the noise is applied and,
contrastingly, refer to this as {\it weak stochastic stability}. 

\subsection{Numerical behaviour of full stochastic Kuramoto system}
\begin{figure}[htb]
\centering
\includegraphics[width=70mm,angle=0]{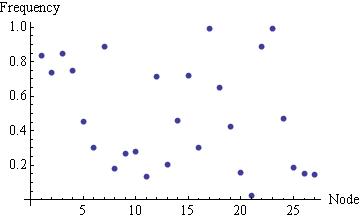}
\caption{Choice of intrinsic frequencies used in this numerical study which is drawn from a
uniform distribution between $[0,1]$} 
\label{FreqDis}
\end{figure}

To give a feel for the complete dynamics, before we proceed with analytical solutions near phase synchrony, we show
the behaviour based on numerical solutions at a value of the coupling where the deterministic system is known to phase synchronise.
We use the RB network \cite{RavBar03} in Fig.\ref{RB-Graph-LaplSpectrum}
for which synchronisation properties of the deterministic
system are known \cite{Kall10}.
We also used in \cite{Kall10} a set of intrinsic frequencies $\omega_i$
drawn from a uniform distribution in the range $\omega_i \in (0,1)$,
for example the set in Fig.\ref{FreqDis}. 
Then the largest values
of the fixed point $\frac{\omega^{(r)}}{\lambda_r}$ are
$\frac{\omega^{(4)}}{\lambda_4}=0.530, \frac{\omega^{(1)}}{\lambda_1}=0.339, 
\frac{\omega^{(3)}}{\lambda_3}=0.241$.  Note that both $r=1,4$ correspond to
eigenvectors with all non-zero components. We therefore examine
the mode, $r=3$, because it involves fewer nodes with non-zero weights (as seen in
Fig.\ref{RBevector}). The
projection of the frequency vector onto this eigenvector is $\omega^{(3)}=0.1843$.
We will use this mode and this frequency projection throughout the paper. 

According to first order considerations of the deterministic system
\cite{Kall10}, for all couplings $\sigma>\frac{\omega^{(4)}}{\lambda_4}=0.53$ 
the system should be stably locked in
a state of phase synchrony. Numerical calculations confirm this \cite{Kall10}. 
We therefore adopt a coupling $\sigma=0.8$, at which we know the
deterministic system is phase synchronised, and
examine the effect of noise on the $r=3$ Laplacian mode. 
Then $x^*=0.3$. The initial configuration is also taken as $x'=0.3$. 
We numerically solve the system Eq.(\ref{fullKura-noise}) using Mathematica for some 1000 instances, with noise as defined in 
Eqs.(\ref{noisefluct1},\ref{noisefluct2}). In the presence of noise we generally find that
$10^6$ points suffice up to $t=20$ time steps to manifest the key dynamics. 
At $\sigma=0.8$ the system synchronises rapidly within a few time
units. Thus $t'=8$ guarantees the system is phase synchronised
before the noise switches on. We choose
\begin{eqnarray}
\Omega_r=1, \  \gamma_1^{(r)} = 1.2 \delta_{r3}, \  \gamma_2^{(r)} = 1.68 \delta_{r3}, \ c_r=0.3 \delta_{r3}. 
\label{noisevals}
\end{eqnarray}
(Note that $\gamma_2=1.4\gamma_1$.) 

At this point we temporarily use the traditional symbol $r$ for Kuramoto's order parameter 
\cite{Kura} (and no longer as a label for the Laplacian modes):
\begin{eqnarray}
r \equiv \frac{1}{N} \left|\sum^N_{j=1} e^{i \theta_j}\right|
\label{orderparam}
\end{eqnarray}
Values of $r\approx 1$ reflect full phase synchrony, $\theta_i=\theta_j \ \forall i,j$. 
At these settings, frequently the noise immediately
causes the order parameter to deviate
from $r=1$ but with a delay
before the entire system of oscillators becomes chaotic.

\begin{figure}[htb]
\centering
\includegraphics[width=60mm,angle=0]{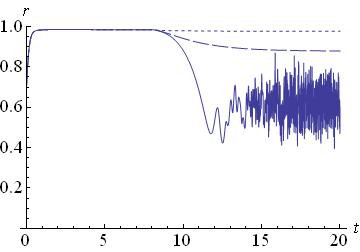}
\caption{\scriptsize{Plot of three instances of the order parameter $r$, Eq.(\ref{orderparam}),
solving for the full Kuramoto model applying noise with
values Eq.(\ref{noisevals}) at $t=8$ to the $\vec{\nu}^{(3)}$ Laplacian eigenvector
over the whole period of the numerical solution up to $t=20$.}} 
\label{unstablerplot}
\end{figure}

At this stage, we set $\kappa_T=20$: the noise is sustained from
the moment it begins to the end of the period over which we solve the system.
We show in Fig.\ref{unstablerplot} three instances of the typical
behaviour. In one there is a delay between $t=8$ and $t\approx 11$ between
the initial impact of the noise and the eventual onset of chaos: $r$ steadily 
drops in value before oscillations begin.
In another instance there is only a slight diminishing of $r$ and stabilisation at some state of 
partial synchronisation.
In others the noise has had little impact over the period for which we solve the system.

\begin{figure}[htb]
\centering
\includegraphics[height=80mm,angle=0]{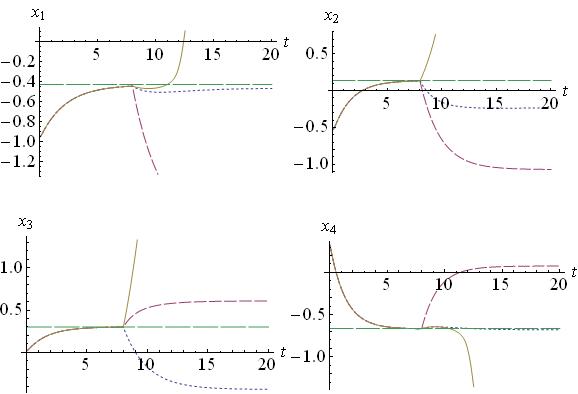}
\caption{\scriptsize{Plots of three instances of the four lowest 
Laplacian modes $x_1,\dots,x_4$ (solid, dotted and short-dashed lines)
applying noise to $\vec{\nu}^{(3)}$ from $t=8$ over the whole period of the 
solution, with dashing corresponding to the instances shown in Fig.\ref{unstablerplot}; the fixed
point values $\frac{\omega}{\sigma\lambda}$ for each of the modes are also indicated (long dashed lines).}} 
\label{unstablexplots}
\end{figure}

Examining the low-lying Laplacian modes $x_1,\dots,x_4$ we can understand this
time delay in the instance where the order parameter becomes chaotic. The modes are plotted in Fig.\ref{unstablexplots}, where the fixed points
$\frac{\omega}{\sigma\lambda}$ for each are also indicated; the modes quickly settle to their fixed point
before the noise is switched on. Because the $\vec{\nu}^{(1)},\dots,\vec{\nu}^{(4)}$ 
eigenvectors all share common nodes with non-zero components there is an immediate 
effect at $t=8$ on all the modes, even though the noise has only been applied
to the third Laplacian eigenvector. However after the noise is switched on each instance behaves quite
differently. For the instance in Fig.\ref{unstablerplot} in which the order parameter becomes chaotic - the solid line - we see that 
in Fig.\ref{unstablexplots} all four of the lowest lying modes
eventually diverge in positive or negative directions; again see the solid line in each case. 
Indeed, the period in which the order parameter decreases monotonically before showing
oscillations coincides with the period in which the mode $x_3$ is {\it approaching} the value one
(demonstrating that there is no `sharp' threshold).
Specifically, for this instance, we observe a difference in behaviour for the modes $x_2, x_3$ against $x_1,x_4$: the latter two remain suppressed for some time before diverging, 
respectively, positively and negatively. This is consistent with the fact, visible in Fig.\ref{RBevector}, that the eigenvectors ${\vec\nu}^{(2)},{\vec\nu}^{(3)}$
share common nodes $i=1,\dots,7$ of the same order of -0.5, while ${\vec\nu}^{(1)},{\vec\nu}^{(4)}$ are somewhat suppressed at these nodes, but non-zero.
However, clearly in Fig.\ref{unstablexplots} the mode $x_3$ reaches the value one fastest, followed by $x_2$, and then modes $x_1$ and $x_4$ turn around. 
We argue that this is evidence for $x_3$ driven to the edge of the basin of attraction first, followed closely by $x_2$, but in the course of this non-linear interations switching on, 
driving $x_1$ and $x_4$ out of steady-state.

For the other instances, dotted and dashed lines in Fig.\ref{unstablexplots}, a type of stabilisation takes place {\it in the presence of noise}
inside the original basin of attraction but not at the deterministic fixed point.
This is evidently a consequence of competition between the deterministic and stochastic aspects of 
the system, about which 
our approach has little to say but warrants investigation in its own right.
For the corresponding instances in Fig.\ref{unstablerplot}, the order parameter has stabilised to some near phase 
synchronisation value.

\begin{figure}[htb]
\centering
\includegraphics[height=50mm,angle=0]{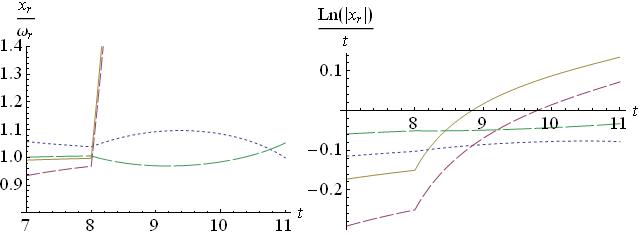}
\caption{\scriptsize{Plots of initial behaviour after switch on of the 
noise for the four lowest Laplacian modes 
for the instance in Fig./\ref{unstablerplot} where the order parameter becomes chaotic; 
on the left $x_r/\omega^{(r)}$ and on the right $\ln|x_r|/t$, with $x_1$ dotted line,
$x_2$ short dashed line, $x_3$ solid line and $x_4$ long dashed line .}}
\label{zoomplot}
\end{figure}

Selecting the particular instance where the order parameter in Fig.\ref{unstablerplot} 
shows chaotic behaviour, 
and zooming in to the behaviour shortly after the noise is switched on we see in 
Fig.\ref{zoomplot} that the 
the deviation away
from the fixed point is essentially linear consistent with the initial behaviour of an exponential. 
In the left hand plot of Fig.\ref{zoomplot}, where we normalise $x_r(t)$ by $\omega_r$, 
we see that when $x_3$ and $x_2$ reach large values then $x_1$ and $x_4$ develop noticeable
curvature. In the right hand plot for $\ln|x_r(t)|/t$ we see that the initial time dependence after the noise begins is
consistent with exponential instability: the exponent is initially linear and has noise dependence, consistent with 
our solution to the Langevin equations, Eq.(\ref{shortlang}). 

The curvature in $x_1$ and $x_4$ in Fig.\ref{zoomplot} coincides with $x_2$ and $x_3$ both of the order
of one: they have left the effective basin of attraction. At this point the variations in the modes
become nonlinear as a consequence of the restored interactions; it is
at this point that the system has left the regime where our approximation
Eq.(\ref{motion2}) applies and the behaviour in Fig.\ref{unstablerplot}
is chaotic. We see that there is a regime in which the linearised system is valid. 
Our aim is to compute an average time over which this regime applies.

\subsection{Fokker-Planck Equations}
Rather than focusing on $x_r$ (where $r$ again labels the mode and not the Kuramoto order parameter) in the Langevin equations
we address the probability
density function for an ensemble of such $x_r$. 
Denoted $P$, it determines the probability
that the random variable $X_r$  lies in the range
$x_r \leq X_r \leq x_r + dx_r$. The Fokker-Planck equation
governs the evolution of this probability density, given the system is initially in
state $X_r=x_r'$ at initial time $t'$:
\begin{eqnarray}
P(\vec{x},t) &\equiv& P(\vec{x},\vec{x}'|t,t')
\nonumber \\
P(\vec{x},t') &=& \delta(\vec{x}-{\vec x}') 
\nonumber 
\end{eqnarray}
where $\vec x$ now represents the vector, in the Laplacian basis, of components $x_r$.
Because we shall always work in time periods where the noise is applied,
we set the Heaviside functions to unity.
From the Langevin equation Eq.(\ref{gora}) we follow Go\'{r}a \cite{2}
(who uses the It\^{o} interpretation) and derive the Fokker Planck equation \cite{Risken},
\begin{equation}
\frac{\partial}{\partial t}P(\vec{x},t) = \sum^{N-1}_{r=1} \frac{\partial^2}{\partial x^2_r}
\left\{ s^{(r)}(x_r) P(\vec{x},t) \right\} - \sum^{N-1}_{r=1} \frac{\partial}{\partial x_r} 
\left \{ q^{(r)}(x_r)  \ P(\vec{x},t) \right\}  
\label{FPY}
\end{equation}
where,
\begin{eqnarray}
s^{(r)}(x_r) &=& \frac{\Omega}{2}
\left\{  \left(\gamma^{(r)}_1 \right)^2 
x^2_r - 2 c_r \gamma^{(r)}_1 \gamma^{(r)}_2 x_r +  \left(\gamma^{(r)}_2 \right)^2   \right\}, \nonumber \\
 q^{(r)}(x_r) &=& \omega^{(r)}  - \sigma \lambda_r x_r.
\label{sfunc}
\end{eqnarray}
Here $s^{(r)}$ represents a diffusion matrix and $q^{(r)}$ a drift vector.

\section{Solving the stationary Fokker-Planck equations and MFPT}
\label{solveFPE}
\subsection{Decoupling}
As the diffusion matrix is diagonal, 
we decompose $P(\vec{x},t)$ into 
a product of $N-1$ normal modes,
\begin{eqnarray}
P(\vec{x},t) = \prod^{N-1}_{r=1}P_r(x_r,t)
\nonumber
\end{eqnarray}
so that Eq.(\ref{FPY}) decouples into separate equations for each $r$
\begin{eqnarray}
  \frac{\partial}{\partial t}P_r(x_r,t) &=&  \frac{\partial^2}{\partial x^2_r}
\left\{ s^{(r)}(x) P_r(x_r,t)  \right\}-  \frac{\partial}{\partial x_r} 
\left \{ q^{(r)}(x)  P_r(x_r,t)  
\right\} 
\label{FPX}\\
 P_r(x_r,t') &=& \delta(x_r -x_r'). \nonumber
\end{eqnarray}
Since we have decoupled the modes we herein drop the 
sub(super)scripts $r$.

\subsection{Stationary densities}
\label{casecase}
As we shall see, the MFPT only relies on the stationary solution of the FP equation, so
we shall not detail the time dependent solution to Eq.(\ref{FPY}). 
Those interested in full time dependence are encouraged to refer to \cite{Linetsky} 
and references therein. 
We solve for the stationary density of Eq.(\ref{FPX}), 
denoted $P_{\rm{st}}(x)$, by setting $\frac{\partial}{\partial t}P(x,t)=0$. We obtain
the corresponding (Pearson's) differential equation,
\begin{eqnarray}
 &&\frac{d}{d x}\left\{ s(x) P_{\rm{st}}(x) \right\} -  q(x)  
\ P_{\rm{st}}(x) =0, \nonumber\\
 \Rightarrow & & P_{\rm{st}}(x) = 
\frac{1}{ s(x)} \exp \left\{  \int^x_{\alpha} dy \frac{q(y)}{s(y)} \right\},
\label{station}
\end{eqnarray}
where the constant $\alpha$ is chosen for convenience. $P_{\rm{st}}$ is 
also referred to as a weight function. 

There are three distinct forms of the stationary densities which are dependent 
on the following choices of $s(x)$:

\begin{itemize}
\item {\it Case I,} $s(x)$ is constant;
\item {\it Case II,} $s(x)$ is quadratic with a single degenerate root; and
\item {\it Case III,} $s(x)$ is quadratic with non degenerate complex roots.
\end{itemize}

We now detail these individual cases. 

\subsubsection{Case I}
For this case we set $\gamma_1 = 0$, hence only additive noise is present 
in the system. We rescale and shift $x$ to
\begin{eqnarray*}
z=\sqrt{\frac{\sigma \lambda}{\Omega \gamma^2_2}}\left(x- \frac{\omega}{\sigma \lambda} \right),
\end{eqnarray*}
in terms of which the normalised stationary probability density function is 
\begin{equation}
P^{(I)}_{\rm{st}}(z) = \sqrt{\frac{\sigma \lambda}{\Omega \gamma^2_2}} e^{ -z^2}.
\label{Hermite-stat}
\end{equation}
The statistical properties of mean, mode and variance are straightforwardly given by
\begin{eqnarray*}
x_{mean} = x_{mode} = \frac{\omega}{\sigma \lambda},&&var 
= \frac{\Omega \gamma^2_2}{2 \sigma \lambda}.
\end{eqnarray*}

Hence this stationary density is a typical Gaussian which has a peak at the mode and 
decays exponentially (symmetrically) on both sides. We note that the mode (namely, the peak value) of 
$P^{(I)}_{\rm{st}}$ is given by the steady state solution of the 
deterministic differential equation (Eq.(\ref{fixedpointdef})).

\subsubsection{Case II}
For this case we set $c=1$, hence this case 
corresponds to maximally correlated noise. Setting the following parameters 
$z$, $\beta_1$ and $\beta_2$ as,
\begin{eqnarray}
z = \gamma_1 x - \gamma_2 ,& \displaystyle \beta_1 =  
-\frac{2 \sigma \lambda}{\Omega \gamma^2_1 } -2,  & \beta_2 = 
\frac{2 \sigma \lambda}{ \Omega \gamma_1  }\left(\frac{\omega}{\sigma \lambda}-  
\frac{\gamma_2}{\gamma_1} \right),
\label{besselbeta-main}
\end{eqnarray}
the normalised stationary distribution is given by,
\begin{equation}
P^{(II)}_{\rm{st}}(z) =\frac{ |\gamma_1|}{\beta^{\beta_1+1}_2}   
\frac{(-\beta_1-1)}{\Gamma\left(-\beta_1 \right) }
z^{\beta_1}e^{-\frac{\beta_2}{z}},
\label{Bessel-stat}
\end{equation}
where $\Gamma(x)$ is the usual `Euler' gamma function.
 
From this it is straightforward to compute properties such as the mean, mode, variance and skewness:
\begin{eqnarray}
x_{mean}  = \frac{\omega}{\sigma \lambda}, \, x_{mode} =  
\frac{\omega + \Omega \gamma_1 \gamma_2}{\sigma \lambda + \Omega \gamma^2_1}, \, 
var = \frac{\left( \frac{\omega}{\sigma \lambda} - \frac{\gamma_2}{\gamma_1} \right)^2}{\frac{2 \sigma \lambda}{\Omega \gamma^2_1}-1}
, \nonumber\\
skew  = 
\frac{2}{\frac{ \sigma \lambda}{\Omega \gamma^2_1}-1}\sqrt{\frac{2 \sigma \lambda}{\Omega \gamma^2_1}-1}.
\label{stat2}
\end{eqnarray}
In particular, we see generally see that $x_{mode}\neq x^*$, meaning that the most probable 
steady state position of $x$ is not the deterministic stationary point. The case $c=-1$ can be generated from these results.

\subsubsection{Case III}
We now constrain $c \ne \pm 1$ and redefine $z$, $\beta_1$ and $\beta_2$ for this case as,
\begin{eqnarray}
z = \frac{\gamma_1 x - c \gamma_2 }{\gamma_2 \sqrt{1-c^2}}, &
\displaystyle \beta_{1} = -\frac{ \sigma \lambda}{\Omega \gamma^2_1}-1, & \beta_2 =  
\frac{2 \left( \frac{\omega}{\sigma \lambda} -  \frac{c\gamma_2}{\gamma_1} \right)}{\frac{\Omega \gamma_1 \gamma_2}{\sigma \lambda}\sqrt{1-c^2}}
\label{romanbeta-main}.
\end{eqnarray}
The normalised stationary density for this process is,
\begin{equation}
P^{(III)}_{\rm{st}}(z)= \frac{|\gamma_1|}{|\gamma_2|\sqrt{1-c^2}}
\frac{(z^2+1)^{\beta_1} e^{ \beta_2 \arctan(z)}}{\int^{\frac{\pi}{2}}_{-\frac{\pi}{2}}
d \theta\frac{ e^{\beta_2 \theta} }{ (cos \theta)^{2(\beta_1+1)} }}.
\label{Roman-stat}
\end{equation}
Key properties for this case include,
\begin{eqnarray}
 x_{mean} =  \frac{\omega}{\sigma \lambda}, \, 
x_{mode} = \frac{\omega + c\Omega \gamma_1 \gamma_2}{\sigma \lambda + \Omega \gamma^2_1}, \, 
var = 
 \frac{\left( \frac{\omega}{\sigma \lambda} - \frac{c\gamma_2}{\gamma_1} \right)^2 +
(1-c^2) \frac{\gamma^2_2}{\gamma^2_1}}{\frac{2 \sigma \lambda}{\Omega \gamma^2_1}-1}, \nonumber\\
skew= \frac{2\left( 1-\frac{c \sigma \lambda \gamma_2}{\omega \gamma_1} \right)}
{\frac{ \sigma \lambda}{\Omega \gamma^2_1}-1}\sqrt{\frac{\frac{2 \sigma \lambda}{\Omega \gamma^2_1}-1}
{1-\frac{2 c \sigma \lambda \gamma_2}{\omega \gamma_1}
+\left( \frac{\sigma \lambda \gamma_2}{\omega \lambda_1} \right)^2}}. \nonumber \\
\label{stat3}
\end{eqnarray}
Carefully taking the limits $c\rightarrow \pm 1, \gamma_1\rightarrow 0$ recovers the results for mean, mode, variance and skewness for the previous two cases; 
however there is more involved in the distribution $P^{(III)}$ itself.
For case III we give examples of parameter settings and corresponding values for properties of the statistical distributions
in Table \ref{choices}.
\begin{table}\begin{center}
\begin{tabular}{| c | c | c | c |c|c|c|}
  \hline                       
 $\frac{\Omega \gamma^2_1}{\omega}$ &$ \frac{\gamma_2}{\gamma_1}$ & $c$&$var$&$x_{mode}$ 
&$MFPT$ \\\hline
 $     0.09$ & $0.58$ & $0.8$ &0.002&0.304 &$4.012 \times 10^{30}$\\
 $0.13$ &  $ 0.6$ & $0.6$ &0.004&0.302& $ 1.438 \times 10^{14}$\\
$0.2$ &$0.8 $ & $0.4$  &0.017 &0.301& $8.552 \times 10^{4} $ \\
 $0.4$ & $-1.1$ & $0.2$ &0.091&0.244&$23.489$\\
 $0.5$ & $-1.7$ & $0.999$ &0.324&0.039& $6.564$\\\hline
\end{tabular}
\caption{Values of the stochastic variables used for the stationary densities in 
Fig.\ref{pic1ROM}, and values of related distributional properties.}
\label{choices}
\end{center}
\end{table}

\begin{figure}[htb]
\centering
\includegraphics[width=100mm,angle=0]{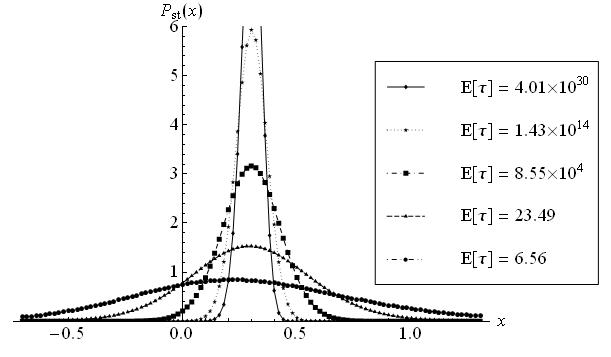}
\caption{\scriptsize{Plots of $P^{(III)}_{st}$ Eq.(\ref{Roman-stat}) with 
$\frac{\omega}{\sigma \lambda}=0.3$. 
Corresponding values of the MFPT, discussed later, and related 
stochastic variables are given in the legend and in Table \ref{choices}.}} 
\label{pic1ROM}
\end{figure}

We give examples of various stationary densities given by Eq.(\ref{Roman-stat})
in Fig.\ref{pic1ROM}, choosing parameters such that $\frac{\omega}{\sigma \lambda} = 0.3$. 

Many of the features of the first and second cases are repeated here, but
with a longer tail of the distribution in the most diffuse case 
(from Eq.(\ref{Roman-stat}) the tail is seen to be
power law suppressed).

\subsection{Computing the MFPT}
The MFPT, denoted by 
\begin{eqnarray}
\mathbb{E}[\tau] \equiv \mathbb{E}[\tau_B| x(t')=x'], \nonumber 
\end{eqnarray}
is the expected value of the time taken for the process in Eq.(\ref{gora}) 
to hit a designated boundary `B', with end points $x=\alpha$ and $\beta$, given that it
begins inside the domain at initial time $t'$: $\alpha < x' < \beta$. 
This is reminiscent of the 
considerations of \cite{intermit} of \textit{on-off intermittency},
the phenomenon of dynamical systems evolving in the vicinity of invariant manifolds 
and undergoing short bursts away from these manifolds. 

The MFPT is calculated from the {\it Andronov-Vitt-Pontryagin} 
equation
\begin{eqnarray}
s(x')\frac{d^2}{dx'^2}\mathbb{E}[\tau] + q(x')\frac{d}{dx'}\mathbb{E}[\tau] = -1,&
& \alpha < x' < \beta,\nonumber \\
\mathbb{E}[\tau|\alpha]=\mathbb{E}[\tau|\beta]=0,
\label{AVP}
\end{eqnarray}
where the initial position is varied keeping the position of the boundary
fixed \cite{Schuss10}. 
The boundary conditions encode that if the system starts on the boundary
it takes zero time to cross it.
The solution to Eq.(\ref{AVP}) is elementary (see \ref{LAPD} for details),
\begin{equation}
\mathbb{E}[\tau] =\frac{\int^{\beta}_{\alpha}\frac{dy}{s(y)P_{st}(y)}
\int^{y}_{\alpha}dz P_{st}(z) }{\int^{\beta}_{\alpha}\frac{dw}{s(w)P_{st}(w)}} 
\int^{x'}_{\alpha}\frac{dv}{s(v)P_{st}(v)} -\int^{x'}_{\alpha}\frac{dy}{s(y)P_{st}(y)}\int^{y}_{\alpha}dz 
P_{st}(z).
\label{MFPT}
\end{equation}

In this paper, we pursue both numerical and, where possible, analytic computations of the MFPT given a solution
to the Fokker-Planck equation. We use the latter to ensure understanding of the behaviour of the former.
For numerical studies, $N$-dependent choices of boundary values
$\alpha,\beta$ are straightforward. Similarly, we shall give formulae through the paper for general choices of
$\alpha,\beta$, though we will concentrate on the minimal choice of $\alpha=-1, \beta=+1$ for the most part.
The legend in Fig.\ref{pic1ROM} provides typical values of MFPT for Case III where this choice of 
$\alpha,\beta$.
We observe that the curve with the highest 
MFPT ($\mathbb{E}[\tau]=4 \times 10^{30}$) 
is, unsurprisingly, also the least diffuse: its variance is very narrow and the even tail is suppressed inside
the region $-1\leq x \leq 1$.
The probability of the system being
near or beyond the effective basin of attraction boundary
is exponentially suppressed. Thus extremely long times
must expire for the system to exit. 
The curve with $\mathbb{E}[\tau]= 1.4 \times 10^{14}$ is very similar. 
Nevertheless, even with small changes in the noise parameters the MFPT has 
decreased by 16 orders of magnitude.
For the remaining plots we can see that the densities are becoming more diffuse and hence their 
MFPT values are quite small: large variances and long tails means that less time is
needed for the system to leave the basin.

As a consequence of the strong peaks of distributions for low MFPT,  we can further approximate the MFPT over times 
for which the decoupled system Eq.(\ref{gora?}) is valid, before nonlinearities dominate the behaviour. 
This is achieved using the Laplace method,
essentially saddle point integration of the integrals in Eq.(\ref{MFPT}).
Recalling that we combined the key deterministic parameters in
$x^{*}=\frac{\omega}{\sigma\lambda}$, 
it is similarly useful to combine stochastic parameters in
\begin{eqnarray}
\mu = \frac{\Omega \gamma^2_1}{\omega}, \ \chi = \frac{\gamma_2}{\gamma_1}. \nonumber
\end{eqnarray}
Setting $\alpha=-1$ and $\beta=1$, we obtain the following approximation of the
MFPT for the most general case ($P^{(III)}_{st}$), for small values:
\begin{eqnarray}
\mathbb{E}^{(III)}_{small}[\tau] \approx  e^{\frac{2 \left(x^*-c \chi \right)}{\sqrt{1-c^2}x^* \mu \chi}\nu_R} 
\left\{ \frac{\left( 1-c^2 \right) \chi^2 + \left( x^* - c \chi \right)^2}{\left( 1-c^2 \right) 
\chi^2 +\left( \frac{x^* - c \chi}{1 + x^* \mu} \right)^2} \right\}^{\frac{1}{x^* \mu}}\nonumber\\
\times \left(1+ \frac{1}{3}\left\{ 1 + x'^2 - 4 x' x^{(III)}_{mode} + 6 \left( x^{(III)}_{mode} \right)^2 
\right\}\frac{|G''_{(III)}(x^{(III)}_{mode})|}{2\Omega}  \right)\nonumber\\
\times \frac{(1-x'^2)  \left(1+x^* \mu \right)^2}{\omega \mu \left\{  (1-c^2)\left(1 + x^* \mu  
\right)^2 \chi^2 + \left(x^*- c \chi \right)^2 \right\}}, \label{laplaceRsmall}
\end{eqnarray}
where,
\begin{eqnarray*}
\frac{|G''_{(III)}(x^{(III)}_{mode})|}{2\Omega} &=& {\frac{ \left( 1 + 
x^* \mu \right)^3}{x^* \mu \left\{ (1-c^2)\left(1 + 
x^* \mu  \right)^2 \chi^2 + \left(x^*- c \chi \right)^2  \right\}}}, 
\end{eqnarray*}
and,
\begin{eqnarray*}
\nu_R = \textrm{arctan}\left( \frac{x^* - c \chi}{\sqrt{1-c^2}\chi \left(1+ x^* \mu \right) } \right) - 
\textrm{arctan}\left( \frac{x^* - c \chi}{\sqrt{1-c^2}\chi  } \right).
\end{eqnarray*}
Details of the derivations of these expressions are given in \ref{LAPD}, where $\alpha,\beta$ are kept general.
This means that an $N$-dependent choice may be implemented in analytical studies for specific graphs,
as discussed earlier. 
The corresponding expression for $P^{(II)}_{st}$ is obtained by taking the $c\rightarrow 1$ limit. The expression for $P^{(I)}_{st}$ is 
then obtained by taking 
$\gamma_1\rightarrow 0$.
In terms of $\mu$ and $\chi$ the corresponding limit is
$\mu \rightarrow 0$, $\chi \rightarrow \infty$ 
such that $\mu \chi^2 \rightarrow \frac{\Omega \gamma^2_2}{\omega}$.

\section{MFPT and Stability: Analytical Results}
We characterise our weaker notion of stochastic stability as 
\begin{equation}
 \mathbb{E}[\tau]  >  \kappa_T,
\label{prescrip}\end{equation} 
namely if the MFPT is greater than some externally selected threshold time, the system is deemed stable.
We therefore restate our earlier question thus: if the system is deterministically
stable, $|x^*|=|\frac{\omega}{\sigma \lambda}|<1$, 
what constraints must be imposed on the combinations of stochastic variables $\mu$ and $\chi$
so that the system also satisfies weak stochastic stability, Eq.(\ref{prescrip})?

Since we are approximating at small MFPT, we can naturally find a condition for small $\kappa_T$. We simply apply 
Eq.(\ref{laplaceRsmall}) to Eq.(\ref{prescrip}) to obtain the inequality,
\begin{eqnarray}
&e^{\frac{2 \left(x^*-c \chi \right)}{\sqrt{1-c^2}x^* \mu \chi}\nu_R} \left( \frac{\left( 1-c^2 \right) \chi^2 + 
\left( x^* - c \chi \right)^2}{\left( 1-c^2 \right) \chi^2 +
\left( \frac{x^* - c \chi}{1 + x^* \mu} \right)^2} \right)^{\frac{1}{x^* \mu}}\nonumber\\
&\times \left\{1 +\frac{(1+x^* \mu) \left[(1+x'^2)(1+x^* \mu)^2 - 4 x' x^* (1+ x^* \mu )(1 + c  \mu \chi)+
6(x^*)^2(1+c \mu \chi)^2  \right]}{3 x^* \mu \left[ (x^* - c \chi)^2+ (1-c^2)\chi^2(1+x^* \mu)^2 \right]}  
\right\} \nonumber\\
&\times \frac{ \left(1+x^* \mu \right)^2}{\mu \left\{  (1-c^2)\left(1 + x^* \mu  \right)^2 \chi^2 + 
\left(x^*- c \chi \right)^2 \right\}}>\frac{ \kappa_T\omega}{(1-x'^2) }.
\label{Romprescipsmall}\end{eqnarray}
Eq.(\ref{Romprescipsmall}) constrains
$\Omega, \gamma_1,\gamma_2$ and $c$ such that, 
given $\omega,\sigma$ and $\lambda$ for deterministic stability and a choice for
the period $\kappa_T$, the system satisfies weak stochastic stability.
We shall illustrate more of its content below with the aid of numerical examples.

\begin{figure}[ht]
\centering
\includegraphics[width=70mm,angle=0]{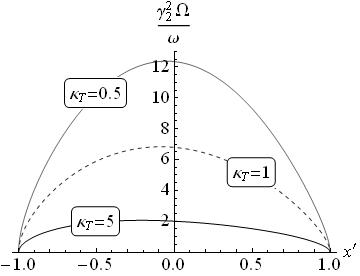}
\caption{Plots of the case where the weak stochastic stability criterion
Eq.(\ref{restic1}) (Case I, $\gamma_1=0$, noise) is an equality, namely equi-MFPT curves,
for various values of the scale factor $\kappa_T$. 
Points above/below the curves represent respectively
 instability/stability. Parameter values are given by $\omega=0.1843$ and  
$x^*=0.3$.}
\label{fig:MFPTHERM}
\end{figure}

Insight into how to unpack Eq.(\ref{Romprescipsmall})
can be obtained by considering the Case I limit
($\mu \rightarrow 0$, $\chi \rightarrow \infty$ 
with $\mu \chi^2 \rightarrow \frac{\Omega \gamma^2_2}{\omega}$). 
For Eq.(\ref{Romprescipsmall}) 
we obtain a quadratic in $\frac{\Omega \gamma^2_2}{\omega}$ given by:
\begin{eqnarray}
\frac{\Omega \gamma^2_2}{\omega} < 
\frac{(1-x'^2)\left(1 + \sqrt{1+\frac{ 4 \kappa_T \omega \{(1+x'^2) + 6  
(x^*)^2 -4x' x^*\}}{ 3x^*(1-x'^2)}}\right)}{2\kappa_T \omega } , 
\label{restic1}\end{eqnarray}
In Eq.(\ref{restic1}) we have discarded solutions which do not satisfy 
deterministic stability, Eq.(\ref{fixedpoint}). In Fig.\ref{fig:MFPTHERM} we plot curves for different $\kappa_T$
where Eq.(\ref{restic1})
is an equality so that regions below/above a given curve represent
weak stochastic stability/instability. The plots are readily understandable: larger values 
of $\kappa_T$ mean a greater amount of time in which the noise can cause the system to drift out of
the basin of attraction; thus lower noise levels, and a smaller area under the curve, are required to maintain stability.
For a given value of $x^*,\ \kappa_T$, one can read thresholds in
$\frac{\Omega\gamma^2_2}{\omega}$ off curves such
as Fig.\ref{fig:MFPTHERM} such that stability is satisfied.

\begin{figure}[htb]
\centering
\includegraphics[width=70mm,angle=0]{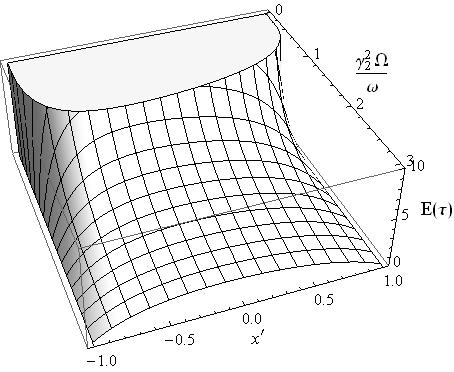}
\caption{\scriptsize{Plot for the exact linearised MFPT for Case I ($\gamma_1=0$) of Eq.(\ref{MFPT}). 
Parameter values are $\omega=0.1843$ and $x^*=0.3$.}} \label{HERMMFPT}
\end{figure}

The approximate level curves in 
Fig.\ref{fig:MFPTHERM} qualitatively emulate the corresponding exact equi-MFPT
lines in Fig.\ref{HERMMFPT},
including asymmetry. When considering Fig.\ref{HERMMFPT}, for any given value of $\mathbb{E}(\tau)$, given by the 
choice of $\kappa_T$, one considers a horizontal plane in the $x'$ and $\frac{\Omega \gamma^2_2}{\omega}$ axes.
Any part of the horizontal plane which is inside its intersection with the surface is considered stable in the linear regime.
Conversely, any part of the horizontal plane which is outside the intersection with the surface is considered unstable in the linear regime. 
The perimeter of the intersection of the surface
 and the horizontal plane gives the lines of equi-MFPT - these are approximately given in Fig.\ref{fig:MFPTHERM} as parabola-like relationships between $x'$ and $\frac{\Omega \gamma^2_2}{\omega}$.

\subsection{Case III MFPT}
We examine the exact MFPT Eq.(\ref{MFPT}) for the noise of the general case III
by plotting it against pairs of variables: MFPT against $x^*$ and $\mu$, and MFPT against $\mu$ and $\chi$. We focus on parameter ranges where 
the MFPT remains small. We consider three values of the correlation $c$ between zero and one.

\begin{figure}[htb]
\centering
\includegraphics[width=130mm,angle=0]{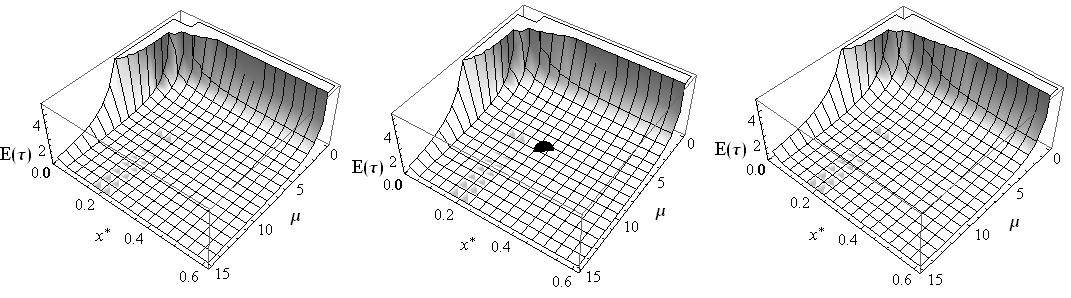}
\caption{\scriptsize{Three plots of the full MFPT Eq.(\ref{MFPT}) against $x^*$ and $\mu$. 
All plots share the parameters $x'=0.3, \omega = 0.1843$ and $\chi=1.4$. 
The left, middle and right plots respectively have the following different $c$ values; 
$c= 0 $, $c= 0.3$ and $c=0.7$. 
The blob in the centre plot represents a choice for which
we earlier numerically solved the full Kuramoto model.}} 
\label{pic-stat1}
\end{figure}
\begin{figure}[htb]
\centering
\includegraphics[width=125mm,angle=0]{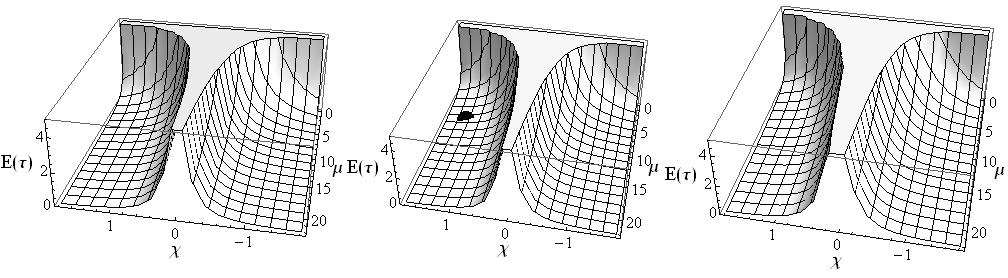}
\caption{\scriptsize{Three plots of the full MFPT Eq.(\ref{MFPT}) against $\mu$ and $\chi$. 
All plots share the parameters $x'=0.3, \omega = 0.1843, 
x^*=0.3$ and the left, centre and right plots respectively 
correspond to $c=0$, $c= 0.3$ and $c=0.7$.
The blob in the centre plot represents a choice for which
the full Kuramoto model was numerically solved earlier.}} 
\label{pic-stat3}
\end{figure}

As with Fig.\ref{fig:MFPTHERM}, when viewing Figs.\ref{pic-stat1} and \ref{pic-stat3} for any given value of $\mathbb{E}(\tau)$, one considers the corresponding linear plane with axes $x^*$ and $\mu$ in Fig.\ref{pic-stat1} and axes $\chi$ and $\mu$ in Fig.\ref{pic-stat3}.
Any part of the plane which is inside its intersection with the surface is considered stable in the linear regime.
Conversely, any part of the plane which is outside its intersection with the surface is considered unstable in the linear regime. 
The perimeter of the intersection of the solid shape and the plane gives the lines of equi-MFPT. These are inverse-like relationships between $x^*$ and $\mu$ in Fig.\ref{pic-stat1}, and inverse-like relationships in $\chi$ and $\mu$ in Fig.\ref{pic-stat3}. The inverse-like relationship in both sets of graphs can be intuitively understood: 
as one of the stochastic parameters becomes large the other must be
reduced to maintain stability.

With such plots one may now read off noise parameter thresholds such that weak stochastic stability is 
satisfied: choosing a threshold in MFPT (vertical axis), the area inside the curve represents stability and the area outside instability,
so that noise parameter ranges can correspondingly be read off.
For example, the `blob' in the centre plots of
Figs.\ref{pic-stat1} and \ref{pic-stat3} indicates
noise parameter values for which we earlier numerically solved the full Kuramoto model.
We see from the plots that for any MFPT greater than one the parameter choice lies outside the region of stability.
This is consistent with how quickly instabilities were typically generated for this noise choice, as seen, for example, in
Figs.\ref{unstablerplot}.

Can we, as for Case I,
gain analytical insight into the curves in Figs.\ref{pic-stat1} and \ref{pic-stat3}?

\subsection{Laplace approximation of Case III MFPT}
\label{analyticst}
To obtain an analytical bound for Eq.(\ref{Romprescipsmall}) we are 
required to make further approximations. It is straightforward to see (for example
from expressions for the variance of the distributions) that small MFPT
values are driven by small values of $\mu$.
Hence for $\mu\rightarrow 0$ we can apply to
Eq.(\ref{Romprescipsmall}) the following simplification:
\begin{eqnarray}
&e^{\frac{2(x^*-c \chi)}{\sqrt{1-c^2}x^* \mu \chi}\nu_R} \left\{ \frac{\left( 1-c^2 \right) \chi^2 + 
\left( x^* - c \chi \right)^2}{\left( 1-c^2 \right) \chi^2 
+\left( \frac{x^* - c \chi}{1 + x^* \mu} \right)^2} \right\}^{\frac{1}{x^* \mu}} \approx 1.
\label{smallsimp}\end{eqnarray}

Inserting Eq.(\ref{smallsimp}) into Eq.(\ref{Romprescipsmall}), we obtain the expression,
\begin{eqnarray}
&\left\{1 +\frac{(1+x^* \mu) \left[(1+x'^2)(1+x^* \mu)^2 - 4 x' x^* (1+ x^* \mu )(1 + c  \mu \chi)+
6(x^*)^2(1+c \mu \chi)^2  \right]}{3 x^* \mu \left[ (x^* - c \chi)^2+ (1-c^2)\chi^2(1+x^* \mu)^2 \right]}  
\right\} \nonumber\\
&\times \frac{ \left(1+x^* \mu \right)^2}{\mu \left\{  (1-c^2)\left(1 + x^* \mu  \right)^2 \chi^2 + 
\left(x^*- c \chi \right)^2 \right\}}>\frac{ \kappa_T\omega}{(1-x'^2) }.
\label{restic3small}
\end{eqnarray}
We notice that the LHS of Eq.(\ref{restic3small}) is quintic in $\mu$ and 
quartic in $\chi$ in contrast to the quadratic dependence on $\Omega \gamma_2^2/\omega$ for case I. 

\begin{figure}[htb]
\centering
\includegraphics[width=120mm,angle=0]{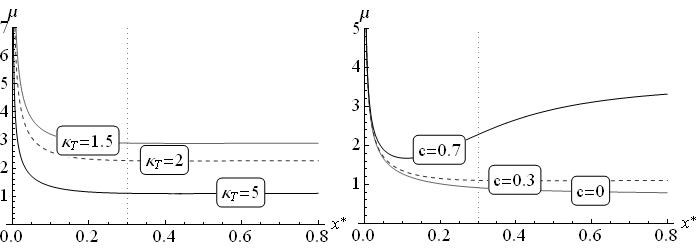}
\caption{\scriptsize{Plots of Eq.(\ref{restic3small}) for $x^*$ against $\mu$ where 
the curves represent lines of 
Equi-MFPT and points lying below and above the curves are respectively classed as 
stable and unstable. All plots on the left hand side have parameter values $\omega=0.1843$, 
$c=0.3$ and $\chi=1.4$ while varying $\kappa_T$. All plots on the right hand side have 
parameter values $\kappa_T=5$, $\omega=0.1843$ and $\chi=1.4$ while varying $c$. }}
\label{ROMcurve3}
\end{figure}

\begin{figure}[htb]
\centering
\includegraphics[width=120mm,angle=0]{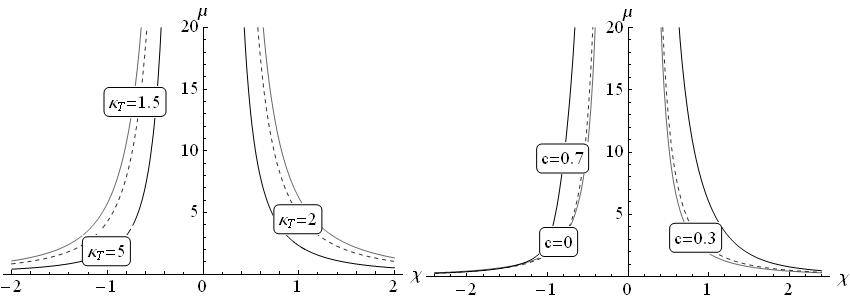}
\caption{\scriptsize{Plots of Eq.(\ref{restic3small}) for $\mu$ against $\chi$, where again 
the curves represent lines of 
Equi-MFPT. Points lying inside and outside the curves are respectively classed as 
stable and unstable. All plots on the left hand side have parameter values $\omega=0.1843$, 
$c=0.3$ and $x^*=0.3$ while varying $\kappa_T$. All plots on the right hand side have parameter 
values $\kappa_T=5$, $\omega=0.1843$ and $x^*=0.3$ while varying $c$.}}
\label{ROMcurve2}
\end{figure}

Obtaining regions of stability from Eq.(\ref{restic3small}) involves 
numerically solving for the roots of $\mu$ and $\chi$. To compare against the plots in 
Figs.\ref{pic-stat1} and \ref{pic-stat3} in the small MFPT regime we plot 
in Figs.\ref{ROMcurve3} and \ref{ROMcurve2}
the points where Eq.(\ref{restic3small}) becomes an equality
against $x^*$ and $\mu$, and against $\mu$ and $\chi$. 
For the most part, the profiles based on the Laplace approximation for small MFPT
qualitatively emulate those of the full MFPT computation in Fig.\ref{pic-stat1}. The exception is
with the large $c$ result - right hand plot, Fig.\ref{ROMcurve3} - with a dip at low $x^*$
for $c=0.7$ which is not reproduced in the exact MFPT plot of Fig.\ref{pic-stat1}. This is readily
understood as a region where the MFPT assumes intermediate values inconsistent with the 
approximations leading to Eq.(\ref{restic3small}). 
Similarly, the plots for small MFPT Fig.\ref{ROMcurve2} reproduce the 
shape of the equi-MFPT profiles of Fig.\ref{pic-stat3} but finer details - 
the peak in Fig.\ref{ROMcurve2} is not as narrow as that in Fig.\ref{pic-stat3} - are lost
for intermediate MFPT values where the underlying approximations breakdown.

Generally then, there are interplays between $\gamma_1, \gamma_2$ and $c$ by which stochastic
stability may be achieved for given $x^*<1$ and $\kappa_T$.
But Eq.(\ref{restic3small}) provide now analytic expressions by which 
approximate thresholds
may be determined, with finer calculation extracted from the full MFPT or refinements
of these methods with higher order terms.

\section{MFPT and Stability: Numerical}
\subsection{A conditional MFPT}
In this section we study the behaviour of the stochastic Kuramoto model comparing switching off
the noise based on our MFPT valid near-phase-synchronisation
with switching off based on an MFPT for the full systeml Eq.(\ref{fullKura-noise}).
To determine an `exact' MFPT we numerically solve Eq.(\ref{fullKura-noise}), 
as discussed earlier, and collect the time for which each instance of $x_3(t)$ first crosses the
threshold $x_3=\pm1$. We then compute the mean over an ensemble of these values of time.
The key issue is that, as said earlier, due to non-linear interactions switching on as the boundary
of the basin of attraction is approached, many configurations in fact reach new stable positions
inside the basin: the values $\pm1$ are never reached. We must therefore collect a {\it conditional} sample:
the time to reach the boundary for the set of configurations that do reach the boundary
within the time for which we solve the system (we could also allow for extrapolations of trajectories
beyond the maximum time for which we solve, which would give a yet longer MFPT).
This is quite different from the MFPT we analytically compute, but gives us something with which
to compare.

\begin{figure}[htb]
\centering
\includegraphics[width=90mm,angle=0]{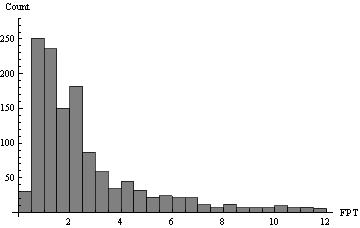}
\caption{\scriptsize{Histogram of values of First Passage Time for the mode $x_3$
to cross the values $\pm1$ when the
noise parameters Eq.(\ref{noisevals}) are applied to $\vec{\nu}^{(3)}$ 
from $t=10$ fo rthe full system Eq.(\ref{fullKura-noise}). The sample is conditioned on the modes reaching the threshold
within the time over which the system is numerically solved. There are 1277 samples in this ensemble with a mean 
of $2.623$. }} 
\label{FPT-histogram}
\end{figure}

Fig.\ref{FPT-histogram} shows a histogram of this conditional sample. Denoting the expected value of the first passage time  
for $x_3$ over this conditioned sample $\mathbb{E}_3^C$, we obtain:
\begin{eqnarray}
\mathbb{E}_3^C=2.623.
\end{eqnarray}
For the same values of noise parameters, the MFPT based on Eq.(\ref{MFPT}) 
(near phase-synchronisation, but no other approximations) gives:
\begin{eqnarray}
\mathbb{E}_3=0.3335.
\end{eqnarray}
The latter is less than the former, again, because non-linearities in the former lead to some trajectories never leaving the
basin of attraction.

\subsection{Switching off the noise}
We now compare the impact of switching off the noise in time $\mathbb{E}_3^C$ versus $\mathbb{E}_3$.
\begin{figure}[htb]
\centering
\includegraphics[width=150mm,angle=0]{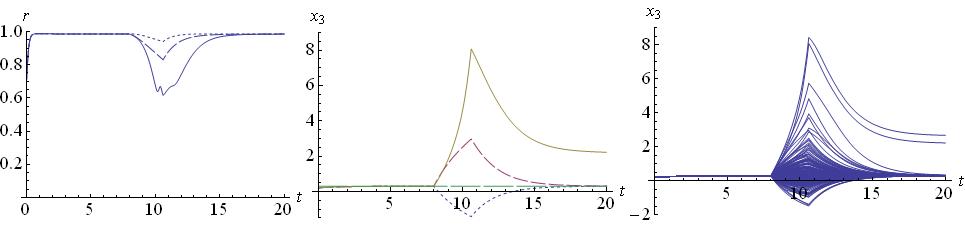}
\caption{\scriptsize{Plots of the behaviour of the full system Eq.(\ref{fullKura-noise}) switching off the noise
based on a conditionally based MFPT, $\mathbb{E}_3^C=2.62$, with
noise parameters Eq.(\ref{noisevals}) applied to $\vec{\nu}^{(3)}$ 
between $10\leq t \leq 12.6$. Left: three instances of the order parameter;
Middle: the behaviour of $x_3$ for the same instances, also showing the 
steady state solution (long dashes); Right: the behaviour of $x_3$
for 150 instances in the numerical solution. }} 
\label{mediumstableplots}
\end{figure}
In Fig.\ref{mediumstableplots} we show the behaviour of the Kuramoto order parameter and the mode $x_3$
when noise is switched off at time $\kappa_T=2.6$, based on $\mathbb{E}_3^C$.
In particular, we see the noise force the system out of the behaviour
where the order parameter is close to one, and drop steadily (left hand plot). 
In most cases, the noise switches off before any chaotic dynamics
emerge, and the deterministic synchronising
dynamics resume, bringing the system back up to order parameter values of one,
and modes $x_3$ back to the fixed point. However there is one instance where
the destabilisation is sufficient that chaotic behaviour shows up in the order parameter
(left hand plot, curve that drops lowest). 
Examining the mode $x_3$ in Fig.\ref{mediumstableplots} for this case (middle plot),
we see that the mode reaches the edge
of the basin - the outward trajectory is very clear - before the noise cuts out and most
of the instances return to the original fixed point. 
Indeed, the instance for which the order parameter has shown an initial signal of chaos in the order parameter (left hand plot)
corresponds to where $x_3$ attains a value very much larger than one (middle plot, highest curve),
before reducing on cessation of the noise; in fact this instance returns to a $2n\pi$ copy of
the original fixed point. Switching off the noise based on  $\mathbb{E}_3^C$ does not
`save' the system from some form of instability: individual modes are deflected
sufficiently from the fixed point basin that they recouple and phase synchronisation
is lost. However, the system always recovers upon cessation of the noise, as seen across a large number of instances
in the right hand plot of Fig.\ref{mediumstableplots}. 

\begin{figure}[htb]
\centering
\includegraphics[width=150mm,angle=0]{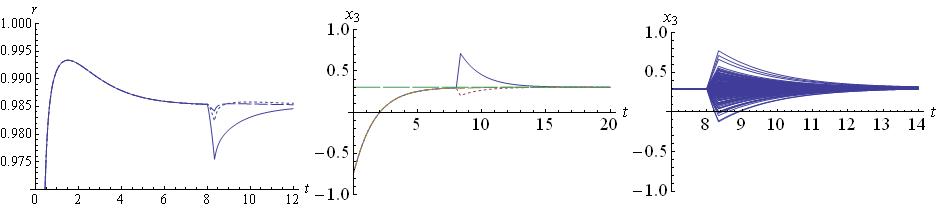}
\caption{\scriptsize{Plots of the behaviour switching off the noise
based on a conditionally based MFPT with
noise parameters Eq.(\ref{noisevals}) applied to $\vec{\nu}^{(3)}$ 
between $10\leq t \leq 10.3335$. Left: three instances of the order parameter;
Middle: the behaviour of $x_3$ for the same instances, also showing the 
steady state solution (long dashes); Right: the behaviour of $x_3$
for 150 instances in the numerical solution. }} 
\label{shortstableplots}
\end{figure}

In contrast, switching off the noise at time $\kappa_T=0.33$, based on the MFPT $\mathbb{E}_3$ from the near-phase-synchronised 
approximation shows no examples of instability. The impact of the noise is so slight that for the order parameter in Fig.\ref{shortstableplots} (left hand plot) we
must zoom into the behaviour near the value one to detect a deviation from full synchrony. Here, the time over which noise is applied is insufficient 
for interactions to switch back on, seen in the modes $x_3$ only slightly deflecting from steady-state values with none reaching values $x_3=\pm 1$.
Alternately, over the period for which noise is applied up to a time based on the MFPT,
the approximation near phase synchrony is valid. 

\section{Summary and Discussion}
\label{discuss}
We have shown that a threshold time for stability of phase synchronisation 
in the general network Kuramoto model subject to
noise can be analytically computed using the Fokker-Planck formalism,
leading to the Mean First Passage Time (MFPT). We have exploited a decomposition in graph
Laplacian modes and a heuristic choice of boundary values of order one for the basin of attraction in
these modes. The noise is applied additively and multiplicatively,
with varying degrees of correlation. An approximation close to the phase synchronised
fixed point has been critical to this. Because non-linear effects, outside our approximation,
lead to unusual stabilisation effects, the MFPT is consistently an under-estimate (compared to, say,
a conditional MFPT based on 
an ensemble of numerical solutions to the exact system that do leave the basin of attraction) of
how long the noise can be sustained before instability is achieved. We gave arguments why for some large $N$
networks the boundary of the basin may be taken to scale with $\sqrt{N}$, and thus there is scope for allowing larger values of MFPT,
but the choice of boundary of order one is sufficient to give a lower bound.
The MFPT in the linearised system can therefore
be used as a threshold for stability, allowing for analytic study of any undirected graph and
frequency distribution. With these results one may determine conservative conditions on noise for stability for
a given network and frequency distribution. Conversely one may use these results to test the robustness of a variety of networks
for given values of noise.

On the basis of our results we can draw some broad implications for the behaviour of a variety
known complex networks.
For deterministic systems on easily disconnected graphs
larger coupling is needed to stabilise phase synchronisation 
due to the ratio $\frac{\omega^{(r)}}{\sigma\lambda_r}$.
With noise we have found that this ratio of deterministic parameters plays off against
the combinations of stochastic parameters $\Omega\gamma_1^2/\omega$ and $\gamma_2/\gamma_1$.
Our basic result is that level curves in MFPT show
an inverse proportionality between these deterministic and stochastic parameters.
This means that the same structural properties that make the system
deterministically stable (lower $\frac{\omega^{(r)}}{\sigma\lambda_r}$)
also favour stochastic stability, thus allowing greater range of $\Omega\gamma_1^2/\omega$ and $\gamma_2/\gamma_1$.
Specifically, for fixed coupling and frequencies, complex networks with 
a bulk of eigenvalues $\lambda_r \ll 1$ 
are both deterministically stable and robust against such a diversity of forms of noise.
Dense Erd{\"o}s-R{\'e}nyi (if generated with
high rewiring probability from a regular graph) and Small World graphs,
have precisely such spectral properties \cite{Jam06}.
Contrastingly, sparse Erd{\"o}s-R{\'e}nyi and Scale-Free graphs
have Laplacian spectra with a `bulk' in the region $\lambda_r \sim 1$
due to a large number of subgraphs disconnectable
by a few link removals. Such graphs are, thus, more susceptible to stochastic
instability. 

This same mechanism may also explain for the deterministic Kuramoto model
the {\it synchronisability} of particular graphs
(how {\it fast} the graph reaches synchronisation rather than its stability
once it gets there); 
for example,
the Small-World graph is known to be more synchronisable \cite{Hong02}.
Close to synchronisation but before onset of the linear regime, most
of the low-lying Laplacian modes may be understood to be nearly quiescent 
generating a bath of
noise on the remaining dynamical Laplacian modes. The synchronisability
problem for the deterministic system may be recast as one of
the system near phase synchronisation subject to noise, analogous
to how a stochastic Kuramoto system can be derived from 
the deterministically chaotic R\"ossler system \cite{Pikov97}; as with
the R\"ossler system, one must make some strong assumptions 
to identify the noise as Gaussian White Noise in this way. 
In this respect, given our result that Small World networks are more stochastically stable, we can
then infer that they are more synchronisable - thus providing a mechanism for the result in \cite{Hong02}

Over and above understanding the impact of noise on well-known complex networks,
for us the power of our analytic results (or even numerically computing
the MFPT from the solution to the Andronov-Vitt-Pontryagin equation) is that one may {\it design} networks
to be robust against noise of specific levels.

To say more about transitions from phase synchronicity to
alternative fixed points or
meta-stable states (such as seen by \cite{ParkKim96,Kim97}) 
requires stochastically probing beyond first order
approximations. For example, in our earlier work \cite{Kall10}
we saw that second order approximations led to the Lotka-Volterra, rather than logistic,
equations. This may help in understanding the non-linear stability behaviours we saw in numerical solutions
of the exact system. It also offers the tantalising possibility of exploiting work on the impact of noise on multi-species
competition dynamics to make more precise conclusions about phase synchronisation. 

\section*{Acknowledgements}
We thank Murray Batchelor, Richard Taylor, Tony Dekker and Markus Brede for 
valuable discussions and comments. 
This work was partly
supported through the Internship program of the Australian
Mathematical Sciences Institute (AMSI) and the Brazilian research agency FAPESP.

\appendix

\section{The incidence matrix and ranges of Laplacian eigenmodes}
We can relate each phase difference $ \theta_j-\theta_k$ to the weighted sum of the eigenmodes $x_r$, using the
construct of the oriented incidence matrix $B$ defined as follows.
Assign an arbitrary direction to each link $a$ of the original unoriented graph $G$.
The matrix elements of the oriented incidence matrix $B_{ia}$
are given by
\begin{eqnarray*}
B_{ia} = \left\{ \begin{array}{ll}
+1 & \textrm{link $a$ is incoming to node $i$}\\
0 & \textrm{link $a$ is not connected to node $i$} \\
-1 & \textrm{link $a$ is outgoing to node $i$}
\end{array} \right. .
\end{eqnarray*}
The Laplacian is then constructed by the following sum over links,
\begin{eqnarray*}
L_{ij} = \sum_{a \in links}B_{ia}B_{aj}.
\end{eqnarray*}
The $BB^T$ structure means that the orientation plays no further role in the Laplacian of the original
unoriented graph or its spectrum.
However, the incidence matrix may be used to transform between
objects respectively in the node and link spaces,
including the Laplacian eigenvectors:
\begin{eqnarray}
e^{(r)}_a = \sum^N_{i=1}B_{ai}\nu^{(r)}_i, &&\sum_{a \in links}e^{(r)}_a e^{(s)}_a = \lambda_r \delta_{rs}.\label{1}
\end{eqnarray}
One can also use the incidence matrix to form a convenient expression for the difference of the phases connected by link $a$ between nodes $j$ and $k$,
\begin{eqnarray}
\Delta \theta_a = \theta_j-\theta_k
= \sum^N_{i=1}B_{ai}\theta_i. \label{phaseDIFF}
\end{eqnarray}
In the above expression we have used the convention that link $a$ is incoming to node $j$ and outgoing to node $k$. Applying $\nu^{(r)}_i$ to both sides of Eq.(\ref{1}) and summing over all $i$ we can solve for $\nu^{(r)}_j$ to obtain,
\begin{eqnarray}
\sum_{a \in links} B_{ja}e^{(r)}_a &=& \sum^N_{i=1}\nu^{(r)}_i L_{ij}, \nonumber \\
\Rightarrow \nu^{(r)}_j &=& \frac{1}{\lambda_r}\sum_{a \in links} B_{ja}e^{(r)}_a.\label{3}
\end{eqnarray}
Hence, beginning with Eq.(\ref{mode-exp}) $(x_r = \sum^N_{i=1}\theta_i \nu^{(r)}_i)$ and Eq.(\ref{phaseDIFF}), we can express the eigenmodes in terms of the phase differences, and the phase differences in terms of the eigenmodes respectively,
\begin{eqnarray}
x_r =  \frac{1}{\lambda_r} \sum_{a \in links} \Delta \theta_a e^{(r)}_a, && \Delta \theta_a = \sum_{r=0}^{N-1}x_r \sum^N_{i=1}B_{ai}\nu^{(r)}_i.\label{4}
\end{eqnarray}

We see that Eq.(\ref{4}) relates the difference of $\theta_j$ and $\theta_k$ with the sum of the difference of the eigenvectors, multiplied by the corresponding eigenmodes. 

As stated earlier, the linear approximation to the sine interaction in the Kuramoto equation only holds when $\left|\Delta \theta_{a}\right| < 1$. To give some solid (albeit simplified) bounds, we consider 
a graph of $N$ nodes consisting of two equally sized complete graphs of order $\sim N/2$ connected by a few bridging links around nodes $i=N/2$.
The Laplacian eigenvalues of this type of graph show marked differences with $\lambda_1 \approx 1$, and $\lambda_r \gg \lambda_1$. 
The corresponding eigenvectors can then be characterised as follows:
\begin{eqnarray*}
\nu^{(1)}_i &=& \left\{ \begin{array}{lll} 
\pm \frac{1}{\sqrt{N}}, &  i = \{1,\dots, \lfloor N/2 \rfloor\}\\
\mp \frac{1}{\sqrt{N}}, & i = \{ \lfloor N/2 +1 \rfloor, \dots, N\}
\end{array}\right. , \label{nu1} \\
\nu^{(r)}_i &=& \left\{ \begin{array}{lll} 
0, & i = \{1,\dots,N\} \ne j_r,k_r\\
\frac{1}{\sqrt{2}}, & i =j_r\\
- \frac{1}{\sqrt{2}}, & i =k_r \; 
\end{array}\right. , \; r\ne 1 \label{nu-high}.
\end{eqnarray*}

We now consider the dynamical situation where the Kuramoto system is close to phase locked synchrony with all but one
of the $x_r$ at steady-state values, $x_r = \frac{\omega^{(r)}}{\sigma \lambda_r}$, and seek the values of $\Delta\theta_a$ if the
one dynamic $x_r$ approaches the value one.
Taking $r=1$ as the dynamic mode, we obtain
\begin{eqnarray}
\Delta \theta_{a} = x_1 \sum^N_{i=1}B_{ai}\nu^{(1)}_i+\sum^{N-1}_{r=2}  \frac{\omega^{(r)}}{\sigma \lambda_r} \sum^N_{i=1}B_{ai}\nu^{(r)}_i.
\label{DELTa}
\end{eqnarray}
The structure of the incidence matrix and eigenvectors leads to
\begin{eqnarray*}
 \sum^N_{i=1}B_{ai}\nu^{(1)}_i &=& \left\{ \begin{array}{ll}
0, & \textrm{if link $a$ connects nodes with the same} \\
&\textrm{eigenvector values}\\
\pm \frac{2}{\sqrt{N}},& \textrm{if link $a$ connects nodes with different}\\
& \textrm{eigenvector values}
\end{array}\right.\\
\sum^N_{i=1}B_{ai}\nu^{(r)}_i &=& \left\{ \begin{array}{ll}
0, & \textrm{if link $a$ connects nodes with eigenvector} \\
&\textrm{values zero}\\
\pm \frac{1}{\sqrt{2}},& \textrm{if link $a$ connects nodes with eigenvector}\\
& \textrm{values zero and $\pm\frac{1}{\sqrt{2}}$}\\
\pm \frac{2}{\sqrt{2}},& \textrm{if link $a$ connects the two nodes with non zero}\\
& \textrm{eigenvector values}
\end{array}\right.
\end{eqnarray*}
where $r > 1$.
Hence
\begin{eqnarray*}
\sum^{N-1}_{r=2}  \frac{\omega^{(r)}}{\sigma \lambda_r} \sum^N_{i=1}B_{ai}\nu^{(r)}_i = \epsilon^{(2,\dots,N)} \ll 1
\end{eqnarray*}
in Eq.(\ref{DELTa}) since $\lambda_r \gg \lambda_1$. We thus obtain
\begin{eqnarray}
&\Delta \theta_{a} = \left\{ \begin{array}{ll}
\pm \frac{2}{\sqrt{N}} x_1 +\epsilon^{(2,\dots,N)} , & \textrm{or}\\
\epsilon^{(2,\dots,N)}
\end{array}\right. \label{phasediff-1}
\end{eqnarray}
We see that $x_1=1$ is consistent with $\Delta\theta_a<1/\sqrt{N}<1$ for all links, or a boundary in phase differences $\Delta\theta_a<1$ implies
a boundary in $x$ of $x_1<\sqrt{N}$. 

Contrastingly, if we consider the situation with one of $x_l$, $l\ne 1$ being dynamic but $x_1$ at steady state, we obtain
\begin{eqnarray*}
\Delta \theta_{a} = \frac{\omega^{(1)}}{\sigma \lambda_1} \sum^N_{i=1}B_{ia}\nu^{(1)}_i + x_l \sum^N_{i=1}B_{ia}\nu^{(l)}_i +\epsilon^{(2,\dots,N,\hat{l})} .
\end{eqnarray*}
Labeling $\epsilon^{(2,\dots,N,\hat{l})}+  \frac{\omega^{(1)}}{\sigma \lambda_1} \sum^N_{i=1}B_{ia}\nu^{(1)}_i = \delta^{(l)}$, we have,
\begin{eqnarray}
\Delta \theta_{a} =\left\{ \begin{array}{ll}
\pm \frac{2}{\sqrt{2}} x_l+ \delta^{(l)}, & \textrm{or} \\
\pm \frac{1}{\sqrt{2}} x_l + \delta^{(l)}, & \textrm{or}\\
\delta^{(l)}
\end{array}\right. \label{phasediff-high}
\end{eqnarray}
We see now that $\Delta\theta_a$ can take values larger than one if $x_l=1$. However, because $-\sigma\lambda_l$ are
the corresponding Lyapunov exponents for these modes they are exponentially suppressed and will not lead to deviation from phase locked
synchrony.

To illustrate these results we compare $\Delta\theta_a$ for the 27 node RB graph, a graph of 100 nodes form by joining two 50 node complete graphs, and one of 200 nodes formed by joining two 100 node complete graphs.

\begin{figure}[htb]
\centering
\includegraphics[width=150mm,angle=0]{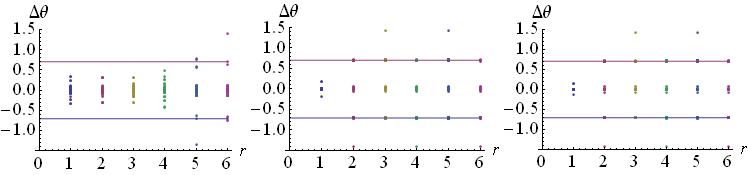}
\caption{\scriptsize{Plot of the scatter of $\Delta\theta$ values across all links in a graph when for a given $r$ the mode $x_r$ is fixed at value one, all others taken
at steady state values $\omega^{(r)}/\sigma\lambda_r$, for three graphs, the 27 node RB graph (left panel), a 100 node graph made from two linked complete 50 node graphs (middle panel), and a 200 node graph made from two linked 100 node complete graphs (right panel). Horizontal lines show the values of $\pm 1/\sqrt{2}$.}} 
\label{DeltaThetaVsR}
\end{figure}
First we show how the values of $\Delta\theta_a$ vary if one of the $x_r=1$ all others taken at steady state values $\omega^{(r)}/\sigma\lambda_r$. 
In Fig.\ref{DeltaThetaVsR}
we show the scatter in $\Delta\theta_a$ across all links $a$ for a given value of $r$ for which $x_r=1$. Note that, rather than generating new frequency
spectra $\omega_i$ for each case here, we simply generate a random sample of values $\omega^{(r)}/\sigma\lambda_r$ but bounded by $\pm 1/\lambda_r$.
We see that in the RB graph (left panel) the modes $r=1,\dots,4$ have $\Delta\theta_a$ values suppressed, while for $r\geq 5$ larger fluctuations 
phase differences can occur. However these are precisely the modes for which $\lambda_r>1$ and therefore have fluctuations exponentially suppressed under
Lyapunov stability. Contrastingly, for $N=100,200$ these larger fluctuations already occur at $r=2$; only $\lambda_1<1$ for these graphs.
We also see the periodicity of these larger fluctuations in units of $1/\sqrt{2}$, consistent with Eq.(\ref{phasediff-high}).
Also in this plot, we see an $N$-dependent suppression of $\Delta\theta$ when the mode reaching value one is $x_1$.
\begin{figure}[htb]
\centering
\includegraphics[width=70mm,angle=0]{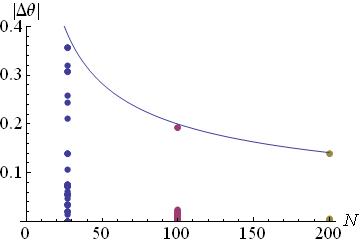}
\caption{\scriptsize{ Plot of the scatter of $|\Delta\theta_a|$ for all links $a$ when $x_1=1$, all other modes taking steady state values, for the three different
$N$ we consider. Superimposed is the curve $1/\sqrt{N}$.}} 
\label{DeltaThetaVsN}
\end{figure}
In Fig.\ref{DeltaThetaVsN} we examine this lowest lying mode case, plotting $|\Delta\theta_a|$ against $N$ for all links $a$. The suppression of $\Delta\theta_a$ by
$1/\sqrt{N}$ is evident.

Finally, at the other extreme, consider a complete graph. This has a completely spectrum of eigenvalues $N-1$ and all eigenvectors take the form
of Eq.(\ref{nu-high}) and the boundary will scale with $\sqrt{2}$. 

\section{Classical stochastic stability analysis of linear stochastic systems}
\label{stabby}
Consider the general linear stochastic system,
\begin{equation}
\dot{x}_l = R_{l}(\vec{x}) + \sum^N_{j=1}S_{lj}(\vec{x})\Gamma_j .
\label{RS}\end{equation}
Following Chapter 11 of \cite{Schuss10}, the classical concept of stability for a stochastic system is 
meaningful only when the fixed point, $\vec{x}^*$, is shared amongst $R_l$ and $S_{lj}$, i.e. 
$R_l(\vec{x}^*) = S_{lj}(\vec{x}^*) = 0$. Due to linearity $\vec{x}^*$ can be scaled to $\vec{0}$ and we rewrite 
Eq.(\ref{RS}) as,
\begin{equation}
\dot{x}_l = \sum^N_{j=1}A_{lj} x_j + \sum^N_{ij=1}\left(B^{(i)}\right)_{lj}x_j \Gamma_i.
\label{formforstab}\end{equation}
A system is classically stochastically stable if, given the matrices $A$ and $B$, 
there exist symmetric positive-definite matrices $Z$ and $C$ satisfying,
\begin{equation}
\sum^N_{j=1}\left( A_{lj}Z_{jm} + Z_{lj}A_{mj} \right) + 
\sum^N_{ijk=1}\left(B^{(i)}\right)_{lj} Z_{jk} \left(B^{(i)}\right)_{mk} = -C_{lm}.
\label{ZandC}\end{equation}
Focusing on the system given by Eq.(\ref{class-stab}) we see that,
\begin{eqnarray*}
A_{ij} = -\sigma \lambda_i \delta_{ij},&\left(B^{(r)}\right)_{ij} = \pm \gamma^{(r)} \delta_{ir} \delta_{jr},
\end{eqnarray*}
and hence Eq.(\ref{ZandC}) becomes,
$$
-\sigma (\lambda_i+\lambda_k) Z_{ik} + \sum^N_{r=1}\left( \gamma^{(r)} \right)^2 Z_{rr} \delta_{ir} \delta_{kr} = - C_{ik},
$$
where we have mapped $r=0$ to $r=N$ for obvious convenience.

Because both $A$ and $B^{(r)}$ are diagonal we may take $Z$ as the $N \times N$ identity matrix. Thus, 
in order for $C$ to be a symmetric positive-definite, and thus for
stochastic stability, we require
$$
\left( \gamma^{(r)} \right)^2 < 2 \sigma \lambda_r.
$$

\section{Approximating the MFPT}
\label{LAPD}
\subsection{Solving the Andronov-Vitt-Pontryagin equation)}
Multiplying both sides of Eq.(\ref{AVP}) by 
$P_{st}(x)$ 
(Eq.(\ref{station})) gives
\begin{equation}
\begin{array}{ll}
&\displaystyle{\frac{d}{dx}\left\{ s(x)P_{st}(x) \frac{d}{dx}\mathbb{E}[\tau] \right\} = - P_{st}(x)}\\
\Rightarrow &\displaystyle{ \mathbb{E}[\tau] = -\int^x_{\alpha}\frac{dy}{s(y)P_{st}(y)}\int^{y}_{\alpha}dz 
P_{st}(z) +\kappa_1 \int^x_{\alpha}\frac{dy}{s(y)P_{st}(y)}+\kappa_2},
\end{array}\label{twoPs}
\end{equation}
where $\kappa_1$ and $\kappa_2$ are integration constants. Substituting $x=\alpha$ 
we obtain $\kappa_2 = 0$, and with $x=\beta$
\begin{eqnarray*}
\kappa_1 = \frac{\int^{\beta}_{\alpha}\frac{dy}{s(y)P_{st}(y)}\int^{y}_{\alpha}dz P_{st}(z) }{\int^{\beta}_{\alpha}
\frac{dw}{s(w)P_{st}(w)}},
\end{eqnarray*}
and hence the full form for $\mathbb{E}[\tau]$ is given by Eq.(\ref{MFPT}).

\subsection{Deriving approximate expressions}
We now show how to obtain the approximation for $\mathbb{E}[\tau]$ 
given by Eq.(\ref{laplaceRsmall}). For this we use the
\textit{Laplace Method}, expected to work well when the arguments of the 
exponentials of the stationary densities have sharp turning points at the modes
of the distributions. We begin at Eq.(\ref{twoPs}) and relabel,
\begin{eqnarray}
&s(x)P_{st}(x) = e^{\frac{F(x)}{\Omega}},&P_{st}(x) = e^{\frac{G(x)}{\Omega}},\nonumber \\
\Rightarrow& G(x) = F(x)-\Omega \log_e(s(x)).
\label{reff}
\end{eqnarray}
We then Taylor expand $F(x)$ and $G(x)$ to second order around $x_F$ and $x_G$ - the points where 
$F'(x_F) = G'(x_G) = 0$. Hence,
$$\begin{array}{lll}
F(x)  \approx  F(x_F) - \frac{|F''(x_F)|}{2}(x-x_F)^2&,&G(x)  \approx  G(x_G) - \frac{|G''(x_G)|}{2}(x-x_G)^2,
\end{array}
$$
where it is important to note that we assume that $\alpha < \{x_F,x_G\} < \beta$. Given the above 
approximations Eq.(\ref{twoPs}) becomes,
\begin{eqnarray}
&\frac{d}{dx}\left\{ e^{-\frac{|F''(x_F)|}{2 \Omega}(x-x_F)^2} \frac{d}{dx}\mathbb{E}[\tau] \right\}
\approx - e^{\frac{G(x_G)-F(x_F)}{\Omega}}e^{-\frac{|G''(x_G)|}{2 \Omega}(x-x_G)^2}\nonumber\\
\Rightarrow &  \frac{d}{dx}\mathbb{E}[\tau] \approx -e^{\frac{G(x_G)-F(x_F)}{\Omega}}\left\{ 
\frac{1}{2}\sqrt{\frac{2 \pi \Omega}{|G''(x_G)|}}\textrm{erf}\left( 
\sqrt{\frac{|G''(x_G)|}{2 \Omega}}(x-x_G) \right)\right.\nonumber\\
& \left. \times e^{\frac{|F''(x_F)|}{2 \Omega}(x-x_F)^2}+ \kappa_1e^{\frac{|F''(x_F)|}{2 \Omega}(x-x_F)^2}  \right\},
\label{lap1}\end{eqnarray}
where $\kappa_1$ is an integration constant and erf is the error function.

We now concentrate on the quadratic exponontial term (expanded around $x_F$) multiplied by the error function (expanded around $x_G$) in Eq.(\ref{lap1}). Eq.(\ref{reff}) assures us that for small $\Omega$, $F(x) \approx G(x)$, and hence for the aforementioned exponential term we make the following further approximation,
$$
\frac{|F''(x_F)|}{2 \Omega}(x-x_F)^2 \approx \frac{|G''(x_G)|}{2 \Omega}(x-x_G)^2.
$$
This allows us to take advantage of the integral,
$$
\int^x_0  \textrm{erf}(y)e^{y^2}dy =\frac{x^2}{\sqrt{\pi}} \,_2F_2\left(\begin{array}{cc}1 & 1\\ \frac{3}{2}&2  \end{array} \left| x^2 \right. \right).
$$
Thus Eq.(\ref{lap1}) becomes,
\begin{eqnarray}
&\mathbb{E}[\tau] \approx -\frac{1}{2}e^{\frac{G(x_G)-F(x_F)}{\Omega}} \left\{ (x-x_G)^2 \,_2F_2\left(\begin{array}{cc}1 & 1\\ \frac{3}{2}&2  \end{array} \left|\frac{|G''(x_G)|}{2 \Omega} (x-x_G)^2 \right. \right) \right.\nonumber\\
&-(\alpha-x_G)^2 \,_2F_2\left(\begin{array}{cc}1 & 1\\ \frac{3}{2}&2  \end{array} \left|\frac{|G''(x_G)|}{2 \Omega} (\alpha-x_G)^2 \right.\right) \nonumber\\
&\left.+ \kappa_1 \left[ \textrm{erfi}\left( \sqrt{\frac{|F''(x_F)|}{2\Omega}}(x-x_F) \right) - \textrm{erfi}\left( \sqrt{\frac{|F''(x_F)|}{2\Omega}}(\alpha-x_F) \right)\right] + \kappa_2 \right\}.\nonumber\\
\label{lap4}
\end{eqnarray}
where erfi is the complex error function. From
$\mathbb{E}[\tau(\alpha)] = \mathbb{E}[\tau(\beta)] = 0$, we obtain $\kappa_2 = 0$ and,
\begin{eqnarray*}
&\kappa_1 = \frac{ (\alpha-x_G)^2 \,_2F_2\left(\begin{array}{cc}1 & 1\\ \frac{3}{2}&2  \end{array} \left|\frac{|G''(x_G)|}{2 \Omega} (\alpha-x_G)^2 \right. \right)}{\textrm{erfi}\left( \sqrt{\frac{|F''(x_F)|}{2\Omega}}(\beta-x_F) \right)-\textrm{erfi}\left( \sqrt{\frac{|F''(x_F)|}{2\Omega}}(\alpha-x_F) \right)}.\\
&-\frac{ (\beta-x_G)^2 \,_2F_2\left(\begin{array}{cc}1 & 1\\ \frac{3}{2}&2  \end{array} \left|\frac{|G''(x_G)|}{2 \Omega} (\beta-x_G)^2 \right. \right)}{\textrm{erfi}\left( \sqrt{\frac{|F''(x_F)|}{2\Omega}}(\beta-x_F) \right)-\textrm{erfi}\left( \sqrt{\frac{|F''(x_F)|}{2\Omega}}(\alpha-x_F) \right)}
\end{eqnarray*}
In Eq.(\ref{lap4}) we apply the asymptotic expansions for the complex error function and the hypergeometric function \cite{obscure},
\begin{eqnarray*}
\textrm{erfi(x)} \thicksim 2 \sqrt{\pi}x ,& \,_2F_2\left(\left.\begin{array}{cc}1 & 1\\ \frac{3}{2}&2  \end{array} \right|x  \right) \thicksim 1+\frac{1}{3}x ,
\end{eqnarray*}
where $x \ll 1$. Hence Eq.(\ref{lap4}) for small $\frac{|F''(x_F)|}{2\Omega}$ and $\frac{|G''(x_G)|}{2\Omega}$ becomes,
\begin{eqnarray}
\mathbb{E}[\tau] \approx \frac{(\beta-x)(x-\alpha)}{2}e^{\frac{G(x_G)-F(x_F)}{\Omega}}\left\{1+ \left[ x^2 + 6 x^2_G + \alpha^2 +\alpha \beta \right.  \right.\nonumber\\
\left. \left. +\beta^2 - 4 x_G (\alpha+\beta) + x (\alpha+\beta-4 x_G)\right] \frac{|G''(x_G)|}{6 \Omega}\right\}.
\label{lapdf}\end{eqnarray}
Eq.(\ref{lapdf}) is the expression used to generate the explicit form of the MFPT in the small valued regime,  Eq.(\ref{laplaceRsmall}). Finally, using 
$(I)$, $(II)$ and $(III)$ to stand for the respective cases (see Section \ref{casecase}), we have explicitly,
\begin{eqnarray*}
&F_{(I)}(x)=& -\frac{\sigma \lambda}{\gamma^2_2}\left( x - \frac{\omega}{\sigma \lambda} \right)^2,\\
&G_{(I)}(x)=& -\frac{\sigma \lambda}{\gamma^2_2}\left( x - \frac{\omega}{\sigma \lambda} \right)^2 +\Omega \log_e \left( \frac{2}{\Omega \gamma^2_2} \right),\\
&F_{(II)}(x)=& -\frac{2 \sigma \lambda}{\gamma_1}\left(\frac{\omega}{\sigma \lambda}-\frac{\gamma_2}{\gamma_1} \right)\frac{1}{\gamma_1 x - \gamma_2} - \frac{2 \sigma \lambda}{\gamma^2_1}\log_e(\gamma_1 x - \gamma_2),\\
&G_{(II)}(x)=& -\frac{2 \sigma \lambda}{\gamma_1}\left(\frac{\omega}{\sigma \lambda}-\frac{\gamma_2}{\gamma_1} \right)\frac{1}{\gamma_1 x - \gamma_2}\\
&& - \left(\frac{2 \sigma \lambda}{\gamma^2_1}+\Omega\right)\log_e(\gamma_1 x - \gamma_2) + \Omega \log_e \left(\frac{2}{\Omega} \right),\\
&F_{(III)}(x)=& \frac{2 \sigma \lambda}{\gamma_1\gamma_2 \sqrt{1-c^2}}\left(\frac{\omega}{\sigma \lambda}-\frac{c \gamma_2}{\gamma_1} \right)\textrm{arctan}\left\{\frac{\gamma_1}{\gamma_2 \sqrt{1-c^2}}\left( x- \frac{c \gamma_2}{\gamma_1} \right) \right\}\\
&& - \frac{ \sigma \lambda}{\gamma^2_1}\log_e \left\{ \frac{\gamma^2_1}{\gamma^2_2(1-c^2)}\left( x - \frac{c \gamma_2}{\gamma_1} \right)^2+1 \right\},\\
&G_{(III)}(x)=& \frac{2 \sigma \lambda}{\gamma_1\gamma_2 \sqrt{1-c^2}}\left(\frac{\omega}{\sigma \lambda}-\frac{c \gamma_2}{\gamma_1} \right)\textrm{arctan}\left\{\frac{\gamma_1}{\gamma_2 \sqrt{1-c^2}}\left( x- \frac{c \gamma_2}{\gamma_1} \right) \right\}\\
&& -\left( \frac{ \sigma \lambda}{\gamma^2_1} + \Omega \right) \log_e \left\{ \frac{\gamma^2_1}{\gamma^2_2(1-c^2)}\left( x - \frac{c \gamma_2}{\gamma_1} \right)^2+1 \right\} \\
&&+  \Omega \log_e \left(\frac{2}{\Omega \gamma^2_2 (1-c^2)} \right),
\end{eqnarray*}
and,
\begin{eqnarray*}
x^{(I)}_F = x^{(I)}_G =  x^{(II)}_F =  x^{(III)}_F =  \frac{\omega}{\sigma \lambda},&
x^{(II)}_G = x^{(II)}_{mode},& x^{(III)}_G = x^{(III)}_{mode}.
\end{eqnarray*}








\end{document}